\documentclass[aps, twocolumn, showpacs, amsmath,superscriptaddress]{revtex4}
\usepackage{amssymb}
\usepackage{dcolumn}
\usepackage{bm}
\usepackage{graphicx}
\usepackage{color}
\usepackage{ulem}
\usepackage{epstopdf}

\DeclareGraphicsExtensions{. jpg,. pdf, . mps, . png, .eps, . ps, . EPS}

\begin{document}
\def\be{\begin{equation}}
\def\ee{\end{equation}}

\def\bc{\begin{center}} 
\def\ec{\end{center}}
\def\bea{\begin{eqnarray}}
\def\eea{\end{eqnarray}}

\newcommand{\avg}[1]{\langle{#1}\rangle}
\newcommand{\ket}[1]{\left |{#1}\right \rangle}
\newcommand{\beq}{\begin{equation}}
\newcommand{\eneq}{\end{equation}}
\newcommand{\beqnn}{\begin{equation*}}
\newcommand{\eneqnn}{\end{equation*}}
\newcommand{\beqy}{\begin{eqnarray}}
\newcommand{\eneqy}{\end{eqnarray}}
\newcommand{\beqynn}{\begin{eqnarray*}}
\newcommand{\eneqynn}{\end{eqnarray*}}
\newcommand{\half}{\mbox{$\textstyle \frac{1}{2}$}}
\newcommand{\proj}[1]{\ket{#1}\bra{#1}}
\newcommand{\av}[1]{\langle #1\rangle}
\newcommand{\braket}[2]{\langle #1 | #2\rangle}
\newcommand{\bra}[1]{\langle #1 | }
\newcommand{\Avg}[1]{\left\langle{#1}\right\rangle}
\newcommand{\inprod}[2]{\braket{#1}{#2}}
\newcommand{\upket}{\ket{\uparrow}}
\newcommand{\downket}{\ket{\downarrow}}
\newcommand{\Tr}{\mathrm{Tr}}
\newcommand{\hcontrol}{*!<0em,.025em>-=-{\Diamond}}
\newcommand{\hctrl}[1]{\hcontrol \qwx[#1] \qw}
\newenvironment{proof}[1][Proof]{\noindent\textbf{#1.} }{\ \rule{0.5em}{0.5em}}
\newtheorem{mytheorem}{Theorem}
\newtheorem{mylemma}{Lemma}
\newtheorem{mycorollary}{Corollary}
\newtheorem{myproposition}{Proposition}
\newcommand{\vp}{\vec{p}}
\newcommand{\Or}{\mathcal{O}}
\newcommand{\so}[1]{{\ignore{#1}}}

\newcommand{\red}[1]{\textcolor{red}{#1}}
\newcommand{\blue}[1]{\textcolor{blue}{#1}}

\title{Complex Quantum Network Geometries: \\Evolution and Phase Transitions }

\author{Ginestra Bianconi}
\affiliation{School of Mathematical Sciences, \\Queen Mary University of London, London E1 4NS, United Kingdom}
\author{Christoph Rahmede}
\affiliation{Institute for Theoretical Physics, \\Karlsruhe Institute of Technology, 76128  Karlsruhe,  Germany}
\author{Zhihao Wu}
\affiliation{School of Computer and Information Technology, Beijing Jiaotong University, Beijing 100044,  China}

\begin{abstract}
Networks are topological and geometric structures used to describe systems as different as the Internet, the brain  or the quantum 
structure of space-time. 
Here we define complex quantum network geometries, describing the underlying structure of growing simplicial 2-complexes, i.e. simplicial complexes formed by  triangles. These networks are 
geometric networks with energies of the links that grow according to a non-equilibrium dynamics. The evolution in time of the geometric  networks is a classical evolution describing a given 
path of a path integral defining the  evolution of  quantum network states. The  quantum network states are characterized by quantum occupation numbers that can be mapped 
respectively to the nodes, links, and triangles incident to each link of the network. We call the geometric networks describing the evolution of quantum network states the  quantum geometric 
networks.
The quantum geometric networks have many properties common to complex networks including small-world property, high clustering coefficient, high modularity, scale-free degree distribution.
Moreover they  can be   distinguished between the Fermi-Dirac Network and the Bose-Einstein Network obeying respectively the Fermi-Dirac and Bose-Einstein 
statistics. We show that these networks can undergo structural phase transitions where the geometrical properties of the networks change drastically.
Finally we comment on the relation between Quantum Complex Network Geometries, spin networks and triangulations.
\end{abstract}


\maketitle
\section{Introduction}

Networks are discrete structures that can be used to describe, model and understand a variety of real systems, including complex interacting systems 
\cite{RMP,Newman_rev,Boccaletti2006,Doro_book,Santo} or the microscopic nature of space-time \cite{spinnet1,spinnet2,Regge,Rovelli}.

Recently, in network science  the characterization of the complexity of networks with geometrical and topological methods is gaining large momentum with several works related to the definition 
of curvature of the 
networks \cite{Yau1,Yau2,Jost,Ollivier,Keller1,Keller2,Higuchi,Knill1,Saniee,Gromov,Mahoney_gromov,Jonck,Bary}, persistent homology \cite{Vaccarino1,Vaccarino2,Mason}, complex networks embedded in finite dimensions \cite{Doro_link,det,Apollonian,Aste2},
and the study of the hyperbolicity of complex networks  
\cite{Kleinberg,Aste,Boguna_navigability,Boguna_Internet,Hyperbolic,Boguna_growing,Clough,Reka,Caldarelli}.

In this context, it  is becoming clear that in order to characterize the geometry of networks it is important to 
describe the underlying structure of simplicial complexes. A simplicial complex is  constructed by gluing together simplices such as points, lines, triangles etc., along their faces. Therefore different works characterize ensembles of random simplicial complexes \cite{Farber1,Farber2,Kahle,Krioukov_SC} and their geometrical and topological properties. Recently a model for 
emergent  complex network geometry has been proposed based on growing simplicial complexes \cite{SR}.

In quantum gravity a central problem is to find the appropriate model describing the geometry of space-time at the quantum level. Different approaches have 
been proposed in which geometry emerges from some pregeometric phase \cite{Wheeler,pregeometry_review,pregeometry2} that include  spin networks and 
loop quantum gravity \cite{spinnet1,spinnet2,Rovelli}, spin foams \cite{Oriti}, causal dynamical triangulations \cite{CDT1,CDT2}, causal sets \cite{Bombelli}, energetic causal sets \cite{Smolin1,Smolin2,Smolin3}, network cosmology \cite{Cosmology}, 
and quantum graphity \cite{graphity_rg,graphity,graphity_c}. Network-like structures play a fundamental role in all these approaches.

This suggests that the emergence of geometric structure from the quantum description of networks is a more general mathematical problem 
that can be not only relevant for understanding the structure of space-time, but which might also help to understand general complex network structures.

Already in the early days of the field of network science the relation between complex network topologies and quantum statistics was shown in the framework of the Bianconi-Barabasi model \cite{Bose,Fitness} that is a growing network model with preferential attachment and 
fitness of the nodes which display a Bose-Einstein condensation. This model can also be extended to   weighted networks \cite{Weighted} described by the Bose-Einstein 
statistics and undergoing also the condensation of the weight of the links. The relation between growing Cayley trees with fitness of the nodes and Fermi-Dirac statistics has been found in \cite{Fermi} and the underlying symmetries between the models in \cite{Bose} and \cite{Fermi} have been discussed in \cite{Complex}.
In the context of equilibrium network models  it has been shown that quantum statistics emerges to describe simple or weighted networks \cite{Garlaschelli}. 

Here we characterize the non-equilibrium  evolution of networks constructed from growing simplicial complexes of dimension two, i.e. formed by triangles, and such that to each link we associate an energy $\epsilon$.
We show that geometrical complex networks emerge from this dynamical evolution which display at the same time  small-world network properties 
\cite{WS}, exponential or scale-free degree distribution \cite{BA}, high clustering coefficient and high modularity \cite{Santo}.
 These networks can be either planar or non-planar with an Euler characteristic that is either $\chi=1$ (planar) or $\chi\propto N$, where $N$ is the 
 network size, indicating a finite average curvature in the network.
 As we will show in two limiting cases of this network dynamics, these networks describe the evolution of quantum network states.
 These network states are constructed along similar lines used in the quantum gravity literature \cite{graphity_rg,graphity,graphity_c} by associating an Hilbert space to each node of the network and two Hilbert spaces to each possible link of the network.
 These network states evolve through a non equilibrium, Markovian dynamics.
The network states can be mapped to geometric networks and  have an evolution described by an appropriate path integral.
The network dynamics describes the paths of single histories of networks on which the path integral is calculated.

We distinguish between Fermi-Dirac Networks and Bose-Einstein Networks. For both of them the number of triangles incident to a given link is $n+1$. However, for Fermi-Dirac Networks $n$ can only
take the values $n=0,1$, whereas for Bose-Einstein Networks $n$ can take any integer value $n=0,1,2\ldots$
These networks evolve in such a way that at the global scale the average of $n$ over the links with energy $\epsilon$ follows  the  Fermi-Dirac statistics for the Fermi-Dirac Network and the Bose-Einstein  statistics for 
the Bose-Einstein Network.

These network structures depend on an external parameter that we call the inverse temperature $\beta$. As a function of $\beta$ they undergo major structural phase transitions in which the network 
structure changes drastically.
In the case of the Fermi-Dirac Network, for $\beta>\beta_c$ the network is not any more small world, but acquires a finite Hausdorff dimensionality.
In the case of the Bose-Einstein Network, for $\beta>\beta_c$ a link acquires a finite fraction of triangles, and the nodes at the end of the link acquire a finite fraction of all the links.
This geometrical phenomenon is the  Bose-Einstein condensation for these networks.

We observe here that the quantum network states studied in this paper are by no means the only way to associate a quantum state  to a network. In particular in the quantum computation community, alternative approaches \cite{Mulken,Huelga,Perseguers,Faccin,Biamonte} have been widely explored, characterizing quantum transport, quantum random networks, and quantum networks in which the links correspond  to entangled states.

The paper is organized as follows. In Sec. II  we describe the  geometric network model with energy of the links depending on the parameter $m$ fixing the maximum number of triangles incident to a link. 
Moreover  we define the entropy rate of the model and we describe the observed phase transition. 
In Sec. III we define the quantum network states.  
In Sec. IV. we define the evolution of the   Fermi-Dirac quantum network state and the Bose-Einstein quantum network state. 
In Sec. V we study the Fermi-Dirac Network (given by the geometric network model with $m=2$). We show that this network characterizes the Fermi-Dirac quantum state, and we show that it is 
globally described by the Fermi-Dirac statistics. Finally we compare the analytical results to simulations and we describe the phase transition occurring at low temperatures.
In Sec. VI we study the Bose-Einstein Network (given by the geometric network model with $m=\infty$) showing that it fully characterizes the Bose-Einstein quantum state, follows 
the Bose-Einstein statistics and undergoes the Bose-Einstein condensation at low temperatures.
In Sec. VII we consider the thermodynamics of the networks and we consider the case in which we project the quantum network state on an unlabeled  final network state.
In Sec. VIII we generalize the geometric model introducing a new parameter $p$ and a new process of addition of triangles that allows for the generation of network geometries that are not planar. 
We characterize the geometry of these networks and we describe the phase transitions observed at low temperature.
In Sec. IX we generalize the evolution of the quantum network states corresponding to the generalized geometric network model.
In Sec. X we describe the dual of the networks generated by the proposed model and we comment on the relation between the Fermi-Dirac Network and spin networks.
In Sec. XI we comment on the relation between Complex Quantum Network Geometries, triangulations and foams.
Finally in Sec. XII we give the conclusions.

\section{Geometric network with energy of the links}
\subsection{Evolution of the geometric networks with energy of the links }
\label{gmb}

Real networks display at the same time several structural properties  (including finite clustering coefficient, significant modularity, finite spectral dimension, heterogeneous  degree distribution) that have been shown to be captured by a very simple model of emergent geometry recently introduced by the authors \cite{SR}.
The model proposed in \cite{SR} is a non-equilibrium model of growing simplicial complexes of dimension $d_n=2$, i.e. formed by gluing triangles along their edges.
In this model  each link can belong at most to a 
 number $m$ of triangles where the parameter $m$ can take any finite value $m\geq 2$ or the value $m=\infty$, indicating the case in which each link can belong to an arbitrarily large number of triangles. In the case $m=2$ the model reproduces random manifolds of dimension $d_n=2$ with an exponential degree distribution and  random distribution of local curvatures, in the case $m=\infty$ the model generates scale-free networks with finite clustering coefficient and significant modularity quantifying the relevance of their community structure.
 
 In  \cite{SR}  all the nodes and all the links are treated equally, having the same probability to attract new triangles. Nevertheless, in complex systems,   attaching a new triangle to a given link might not have the same  probability of  attaching it to another link.

Already in the context of   complex networks growing by preferential attachment \cite{RMP,Newman_rev}, the heterogeneity of the nodes in attracting new links has been recognized to be essential to characterize the evolution of networks, as for example the World-Wide-Web or the Internet \cite{Fitness,Bose}. Usually, this heterogeneous "{quality}" of the nodes is modeled by associating each node to an {\it energy} drawn from a given distribution.
Interestingly, complex networks with energy of the nodes have been shown \cite{Bose,Fermi} to be characterized by quantum Bose-Einstein and Fermi-Dirac statistics, and might display a Bose-Einstein condensation in which one node grabs a finite fraction of the links. This phase transition is relevant for a number of complex networks including economical, technological and social networks in which nodes connected to a finite fraction of the nodes might emerge. 

In the quantum gravity literature, the relation between networks and quantum states has been recently explored \cite{graphity_rg,graphity,graphity_c} to construct   models of emergent space-time geometry. In these works, each network is associated to a quantum network state and the network structure is dictated by  an equilibrium Hamiltonian dynamics.

 Here we consider networks constructed by a non-equilibrium dynamics describing the underlying structure of simplicial complexes constructed  by the addition of connected complexes of dimension $d_n=2$,  i.e. triangles. 
 These networks display non trivial geometrical properties, characterizing in some limit planar random manifolds, as will be discussed later in the paper. For this reason we call them geometric networks.
As in  the geometric network model \cite{SR} we assume that each link can belong at most to a 
 number $m$ of triangles.  Moreover we associate energies both to nodes and links describing the different ability of nodes and links to attract new triangles.
 
 In studying this model our goal is two-fold. On one side we aim at characterizing a wider class of  emergent geometries, and their possible structural phase transitions in order to unveil   the basic geometric properties of complex networks. On the other side we aim at furthering our understanding on the relation between the network evolution and  quantum mechanics by exploring the connection between network evolution, quantum statistics,  and evolution of quantum network states constructed using methods similar to the one introduced in \cite{graphity_rg,graphity,graphity_c}.

The energies of the nodes and of the links are defined as follows. Every node $i$ of the network is associated with the {\it   energy of the link} $\omega_i>0$  drawn from a distribution $g(\omega)$ \cite{note}. The energy $\omega_i$ is assigned to  the node $i$ when the node is added to the network and is quenched during the growth of the network.
Every  link $\ell=(i,j)$  between node $i$ and node $j$ is associated with  the {\it energy of the link} $\epsilon_{ij}$ which is a given symmetric function of the energy of the two nodes $i$ and $j$, i.e. 
\bea
\epsilon_{ij}=f(\omega_i,\omega_j)=f(\omega_j,\omega_i) 
\label{unob}\eea
with $\epsilon_{ij}>0$.

We define the so called {\it spin $J_{ij}$ of the link} ${\ell}=(i,j)$
as 
\bea
J_{ij}=\frac{1}{2}(\omega_i+\omega_j). 
\label{spin}
\eea
The spins of  the links belonging to a triangle between the nodes $i$, $j$ and $r$   satisfy the conditions 
\bea
|J_{ir}-J_{jr}|\leq J_{ij}\leq J_{ir}+J_{jr}.
\label{triangulard}
\eea 
This result remains valid for any permutation of the order of the nodes $i,j$ and $r$ belonging to the triangle.
Although most of the derivations shown in this paper  can be performed similarly for either continuous or discrete energy of the nodes and of the links, here we consider  the case in which the energies of the nodes $\{\omega_i\}$ and the energy of the links $\{\epsilon_{ij}\}$ are discrete. In particular, if the energy of the nodes takes integer values, the spin of the links takes half-integer values and   Eqs. $(\ref{triangulard})$ can be interpreted as the Clebsch-Gordon 
relations between the half-integer spins of the links of each triangle. This property motivates dubbing this variable a spin.

Specific expressions of the energy  $\epsilon_{ij}$ of the link $(i,j)$ might depend  on the spin  $J_{ij}$ of the link.
Examples of  specific choices for the energy of the link are   the quadratic relation,
\bea
\epsilon_{ij}=J_{ij}(J_{ij}+1)
\eea 
or the linear relation 
\bea\label{linearenergy}
\epsilon_{ij}=2J_{ij}=\omega_i+\omega_j. 
\eea
Here we want to keep the generality of the model and we will take $\epsilon_{ij}$ given by Eq. $(\ref{unob})$ unless a specific functional form of the energy of the link is indicated.

The geometric network model is the underlying network of a simplicial complex of dimension $d_2=2$ formed by gluing triangles along the edges.
We assume that each link can belong at most to a number $m$ of triangles where the parameter $m$ can take any finite value $m\geq 2$ or the value $m=\infty$, indicating the case in which each link can belong to an arbitrarily large number of triangles. We call the links to which we can still add at least one triangle unsaturated. All the other links we call saturated. In the case $m=\infty$, all the links are unsaturated.
We start at time $t=1$ from a network formed by a single triangle,  a simplex of dimension $d_n=2$. At each time  we add a triangle to an unsaturated link $(i, j)$ of the network. 
We choose this link  with   probability  $\Pi_{(i, j)}^{[1]}$ given by 
\bea
\Pi^{[1]}_{(i, j)}=\frac{e^{-\beta \epsilon_{ij}}a_{ij}\xi_{ij}(1+n_{ij})}{Z},
\label{prob}
\eea
where $Z=Z$ is given by 
\bea
Z=\sum_{r<s}e^{-\beta\epsilon_{rs}}a_{rs}\xi_{rs}(1+n_{rs}).
\label{Z}
\eea
Here we introduced several time-dependent quantities which we will use repeatedly in this paper:
$a_{ij}$ is the element $(i, j)$ of the adjacency matrix ${\bf a}$ of the network, 
$\xi_{ij}$ is equal to one (i.e. $\xi_{ij}=1$) if the number of triangles to which  
the link $(i, j)$ belongs is less than $m$,  otherwise it is zero (i.e. $\xi_{ij}=0$),
 $n_{ij}+1$ is equal to the total number of triangles incident to the link $(i,j)$.
Having chosen the link $(i, j)$ the simplicial complex at time $t$ is constructed by adding a  node $r$,  two links $(i, r)$ and $(j, r)$ and the new triangle linking node $i$,  node $j$ and node $r$. The geometric complex network is the network structure of the resulting simplicial complex.

Therefore the number of nodes $N$ of the network grows linearly with time and is given by $N=t+2$.

The linking probability depends on the parameter $\beta\geq 0$ that we call {\it inverse temperature}.
For $\beta=0$, all the links that are unsaturated have equal probability to be selected. For  $\beta>0$ instead, unsaturated links with low energy $\epsilon_{ij}$ are more likely to be selected than links with higher energy.
 
\begin{figure}
	\includegraphics[width=0.95\columnwidth]{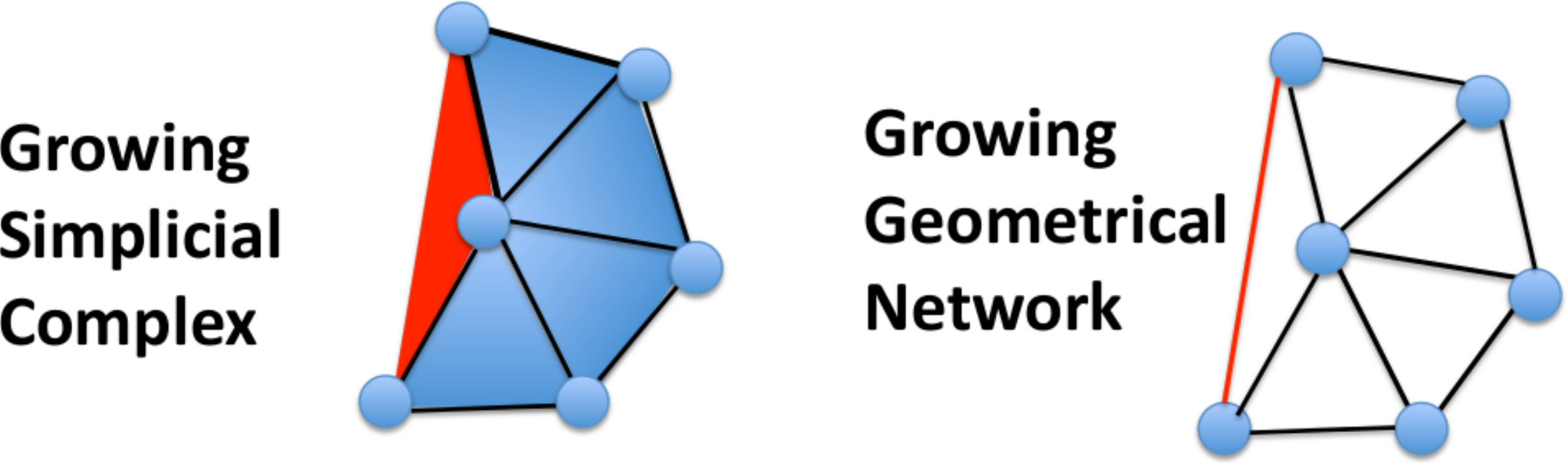}\nonumber \\
	\caption{(Color online)The growing geometrical network is the underlying network structure of a growing simplicial complex in which triangles are continuously attached to the simplicial complex and glued to one unsaturated link. 
	The link where the new triangle is added is chosen with probability $\Pi^{[1]}_{(i,j)}$ given by Eq. $(\ref{prob})$. The figure shows an example where the maximum number of triangles incident to a link is $m=2$. }	
	\label{figuresim1}
\end{figure}

\begin{figure}
$\begin{array}{c}
	\includegraphics[width=0.95\columnwidth]{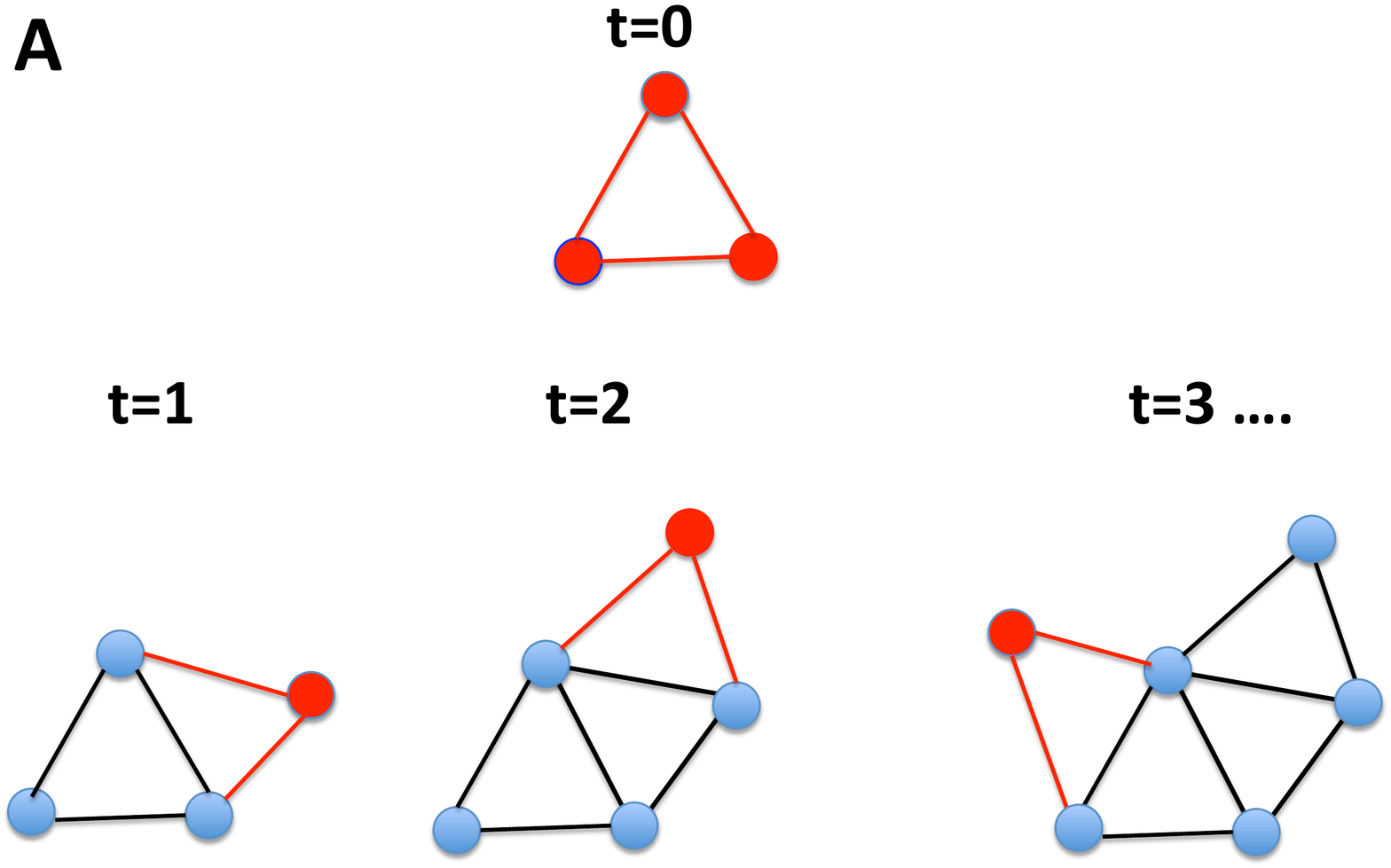}\nonumber \\
	\includegraphics[width=0.95\columnwidth]{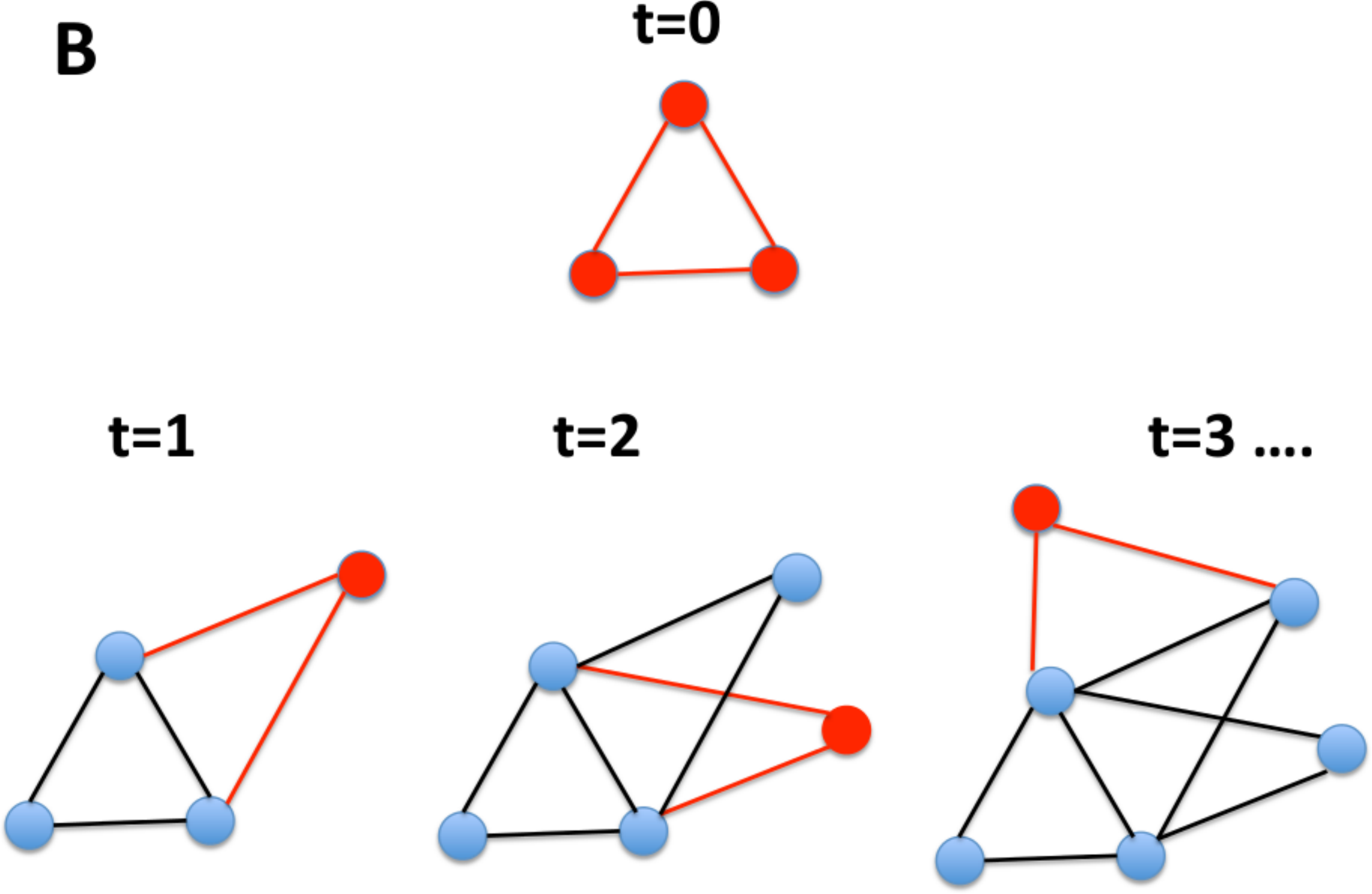}\nonumber \\
	\end{array}$
	\caption{(Color online) The Fermi-Dirac Network evolution (panel A) and the Bose-Einstein Network evolution (panel B). At each time a new triangle is added to a link $(i,j)$ chosen according to the probability $\Pi^{[1]}_{(i, j)}$ given by 
	Eq. $(\ref{prob})$. The maximum number of triangles incident to a link is $m=2$ for the Fermi-Dirac Network evolution and $m=\infty$ for the Bose-Einstein Network evolution. }	
	\label{figure12a}
\end{figure}

With the above algorithm  we describe a growing simplicial complex formed by adding triangles. From this structure we can extract the corresponding network where we consider only the information about node connectivity (which node is linked to which other node).
We call this network model the geometrical growing network.
In Figure \ref{figuresim1} we show schematically the dynamical rules for building the growing simplicial complexes and the growing geometrical  networks that describe its underlying network structure.
In the following we will focus in particular on the limiting cases in which $m=2$ ($2$-dimensional manifolds), or $m=\infty$.
For reasons that  will become clear in the following, we will  call the growing geometric network  with $m=2$ the {\it Fermi-Dirac Network}  and the one with $m=\infty$ the {\it Bose-Einstein Network}.
In Figure $\ref{figure12a}$ we show examples of the first few steps of their evolution.
The Fermi-Dirac Network and the Bose-Einstein Network will also be indicated as quantum geometric networks.
\subsection{Entropy rate of the network evolution}
\label{er}
Entropy measures for network evolution are very important characteristics for evaluating the interplay between randomness and order in these structures \cite{Entropyrate,VonNeumann,Garnerone,Hancock}.
In particular, for growing network models, the entropy rate \cite{Entropyrate} characterizes how the space of typical network dynamical evolutions increases with time.
A change in the scaling of the entropy rate typically indicates a phase transition in the network \cite{Entropyrate}.
The geometric network evolution is described by the sequence $\{\omega(t'),\ell(t')\}_{t'\leq t}$ where $\omega(t)$ indicates the energy of the node added to the network at time $t$, 
$\ell(t)=(i,j)$ indicates the link chosen at time $t$ with probability $\Pi^{[1]}_{(i,j)}$ given by Eq. ($\ref{prob}$). At any given time, therefore, $\ell(t)$ indicates the link to 
which the new triangle is attached.


The entropy rate of the network evolution can be expressed as 
\bea
H_G(t)&=&-\sum_{\omega(t),\ell(t)}P(\omega(t), \ell(t)|\{\omega(t'),\ell(t')\}_{t'< t})\nonumber \\
&&\times\ln P(\omega(t), \ell(t)|\{\omega(t'),\ell(t')\}_{t'< t})
\label{HG}
\eea
where $P(\omega(t), \ell(t)|\{\omega(t'),\ell(t')\}_{t'< t})$ is the probability that, given the temporal evolution  of the network until time $t-1$, at time $t$ a new triangle is attached to the link $\ell(t)$ with   the new node of this triangle having  energy $\omega(t)$. 

At time $t$ the probability that the new node has energy $\omega(t)=\omega$ is given by  the probability distribution $g(\omega)$, and it is independent of the previous evolution of the network. The probability of choosing the link 
$\ell(t)$  is  given by $\Pi^{[1]}_{\ell(t)}$ that depends on the previous history of the network. Moreover $\omega(t)$ and $\ell(t)$ are independent. 
Therefore  the entropy rate can be written as 
\bea
H_{G}(t)=H_{\omega}+H^{[1]}(t)
\eea
with $H_{\omega}$, $H^{[1]}(t)$  specified below.
In particular  $H_{\omega}$ is the contribution to the  entropy rate due to the random distribution of the energy of the nodes, it is independent of time and is given by 
\bea
H_{\omega}=-\sum_{\omega} g(\omega)\ln g(\omega).
\eea
The quantity  $H^{[1]}(t)$ of the  growing geometric network evolution defines the contribution to the entropy rate due to the choice of the link where the new triangle is attached and is given by 
\bea
\hspace*{-9mm}H^{[1]}(t)&=&-\sum_{i<j}\Pi^{[1]}_{(i,j)}\ln\Pi^{[1]}_{(i,j)}
\label{H1}
\eea
evaluated at time $t$.
With Eq. $(\ref{prob})$, we get 
\bea
H^{[1]}(t)&=&\beta\Avg{\epsilon_{ij}}+\log{Z}
\label{H1eZ}
\eea
where 
\bea
\hspace{-5mm}\Avg{\epsilon_{ij}}=\Avg{\epsilon_{ij}\Pi^{[1]}_{ij}}=\sum_{i<j}\frac{a_{ij}\xi_{ij}(1+n_{ij})\epsilon_{ij}e^{-\beta\epsilon_{ij}}}{Z}.
\label{ep}
\eea
We note here that, as the inverse temperature $\beta$ changes, we might expect a phase transition in the network characterized by a different scaling of the entropy rate $H^{[1]}$ and
the normalization constant $Z$ with time $t$ below and above the transition.

The normalization constant $Z$ is fixed by Eq. $(\ref{Z})$.
For $\beta=0$, $Z$ grows linearly with time $t$. In fact for $\beta=0$ and   $m=2$, $n_{ij}=0$ only if $\xi_{ij}=1$ and $n_{ij}=1$ only if $\xi_{ij}=0$. 
Moreover, since at each time we add two unsaturated links and we remove one unsaturated link,
\bea
Z=\sum_{i<j}a_{ij}\xi_{ij}(1+n_{ij})=t+2.
\eea 
For $\beta=0$ and $m=\infty$ instead, all the links are unsaturated, i.e. $\xi_{ij}=1$ and every triangle is incident to three links. Therefore, since we add a triangle for every time step,  
\bea
Z=\sum_{i<j}a_{ij}\xi_{ij}(1+n_{ij})=3t.
\eea 
For a significant range of values of $\beta>0$ we will still continue to have $Z\propto t$ for $t\gg1$ because $Z$ is a sum over a linearly growing set of  non-zero variables.
For $\beta\to \infty$ however, only the unsaturated links with minimal energy of the links $\epsilon_{ij}=\epsilon_0$ will contribute to the sum defined in Eq. $(\ref{Z})$ because the dynamics becomes extremal. Therefore we can have  $Z\simeq {\cal O}(1)$.
In this case a phase transition is expected to occur in the network  at the structural level.
This phase transition induces a substantial change in the geometry of the networks above and below the transition as discussed in the next paragraph.

\subsection{Phase transition in geometric networks}
\label{PT}
The networks constructed according to the model defined in Sec. \ref{gmb} are planar.
In fact the Euler number  of the simplicial complex from which they are extracted is constant during the network evolution and is given by 
\bea
\chi=N-L+T=1
\eea
where $N$ is the total number of nodes, $L$ is the total number of links and $T$ is the total number of triangles.

Let us prove this result recursively.
At time $t=1$ the geometrical networks are formed by a single triangle $N=3,L=3,T=1$.
Therefore we have 
\bea
\chi=1.
\eea
At each time step we add a single node, two links and one triangle, therefore 
\bea
\Delta \chi=\chi(t)-\chi(t-1)=0.
\eea
This shows that the simplicial complexes constructed by gluing new triangles to a single existing link of the network have Euler characteristic $\chi=1$.
When considering the network underlying each of these simplicial complexes, and embedding it in a plane, one can see that the number of faces $F$ of the embedded graph is in fact equal to the number of triangles $T$ of the simplicial complex when we  do not count the external face of the planar network. Therefore these networks are planar. 

Another, equivalent way to prove that our networks are planar is to observe that these networks, by construction, do not contain any complete graphs of five nodes (subgraph $K_5$) or any bipartite complete graph of six nodes (subgraph $K_{3,3}$).

Additionally we define the boundary  of the network, as the set of unsaturated  nodes and links.
Unsaturated links, are links $(i,j)$ with $\xi_{ij}=0$, while unsaturated nodes are nodes with at least an incident unsaturated link.
Note that in the geometric networks studied here, all the nodes are unsaturated since by construction they are always incident to exactly two unsaturated links.
Therefore all the nodes of the network belong to its boundary.  
For these networks the  curvature $R_i$ \cite{Keller1,Keller2} associated to each node $i$ is given by  
\bea\label{networkcurvature}
R_i=1-\frac{k_i}{2}+\frac{T_i}{3}=\frac{4-k_i}{6}=\frac{3-T_i}{6}
\label{Ru}
\eea
where $k_i$ indicates the degree of node $i$, $T_i$ indicates the total number of triangles incident to node $i$, 
and where the last equation can be derived by considering that in the present model $T_i=k_i-1$ for every $i$.
The last expression in Eq. $(\ref{Ru})$, relating the curvature $R_i$ of node $i$ to the number $T_i$ of triangles incident to it, has an intuitive explanation.
As all triangles are isosceles, and each node is at the boundary of the network,each node incident to exactly $T_i=3$ triangles will have zero curvature, since the sum of the angles incident to  it is $\pi$.
 
Here we focus  on the quantum geometric networks (cases $m=2$ and $m=\infty$) and we study the geometry of these network models as a function of $\beta$. These networks are generated by a non-equilibrium dynamics that does not contain any indication about any embedding space.
In the case $m=2$, the Fermi-Dirac Neworks are planar manifolds describing random geometries, 
In the case $m=\infty$, the Bose-Einstein Networks are planar scale-free networks but are not manifolds.

Here  we show numerical evidence that for given distribution $g(\omega)$ and energy of the links $\epsilon_{ij}=f(\omega_i,\omega_j)$
a structural phase transition can occur in quantum geometric networks.
Specifically, we consider the case  in which $\omega$ can only take integer values and  the distribution $g(\omega)$ is Poisson with average $c$, i.e.
\bea\label{Poissondistribution}
g(\omega)=\frac{1}{\omega!}c^{\omega}e^{-c}.
\label{go1}
\eea
Moreover we take the energy $\epsilon_{ij}$ of the generic link $(i,j)$ given by Eq. (\ref{linearenergy}).

\begin{figure*}
	\includegraphics[width=1.9\columnwidth]{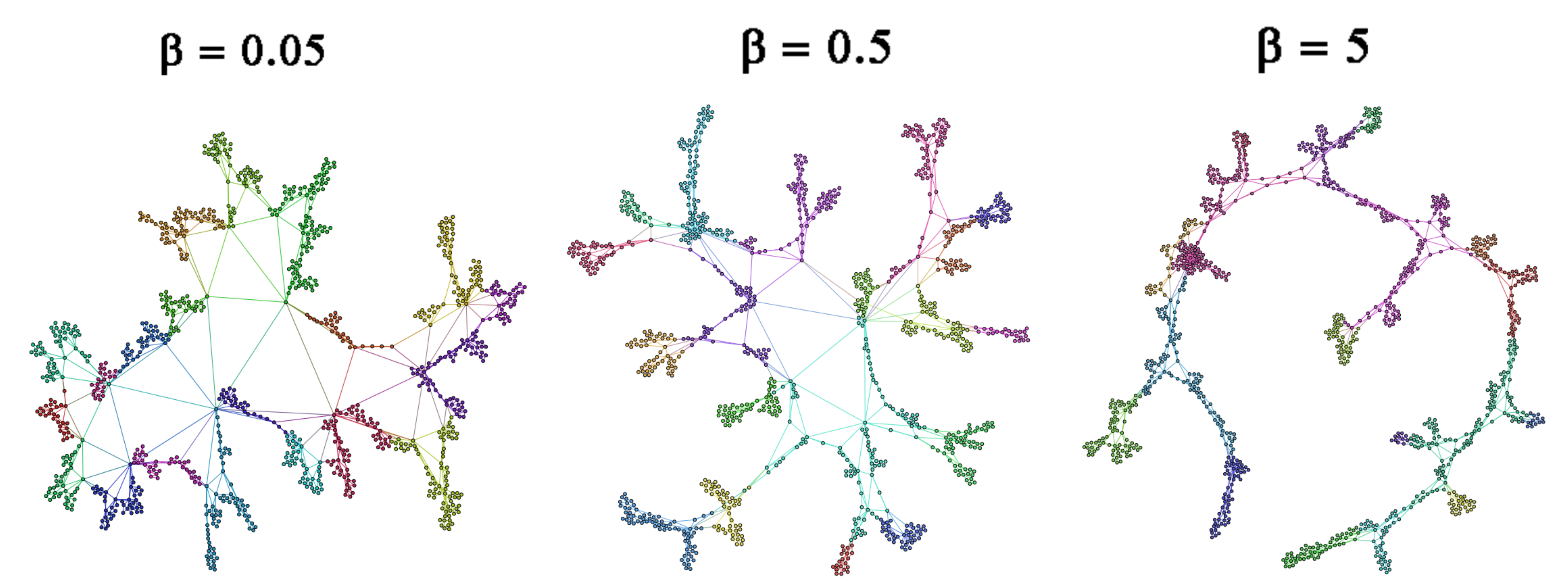}
	\caption{(Color online)Visualization of the  Fermi-Dirac Network  (with  $g(\omega)$ given by Eq. $(\ref{go1})$   and $c=10$)  for $\beta=0.05,0.5,5$ and $N=1000$. For low value of $\beta$, 
	i.e. $\beta<\beta_c\simeq 0.14$, the network is small-world, for large values of $\beta$, i. e. $\beta>\beta_c\simeq 0.14$, the network develops a large diameter. The colour indicates the partition into communities found by running the Louvain algorithm \cite{Louvain}.}	
	\label{visualization_m2}
\end{figure*}

\begin{figure*}
\includegraphics[width=1.9\columnwidth]{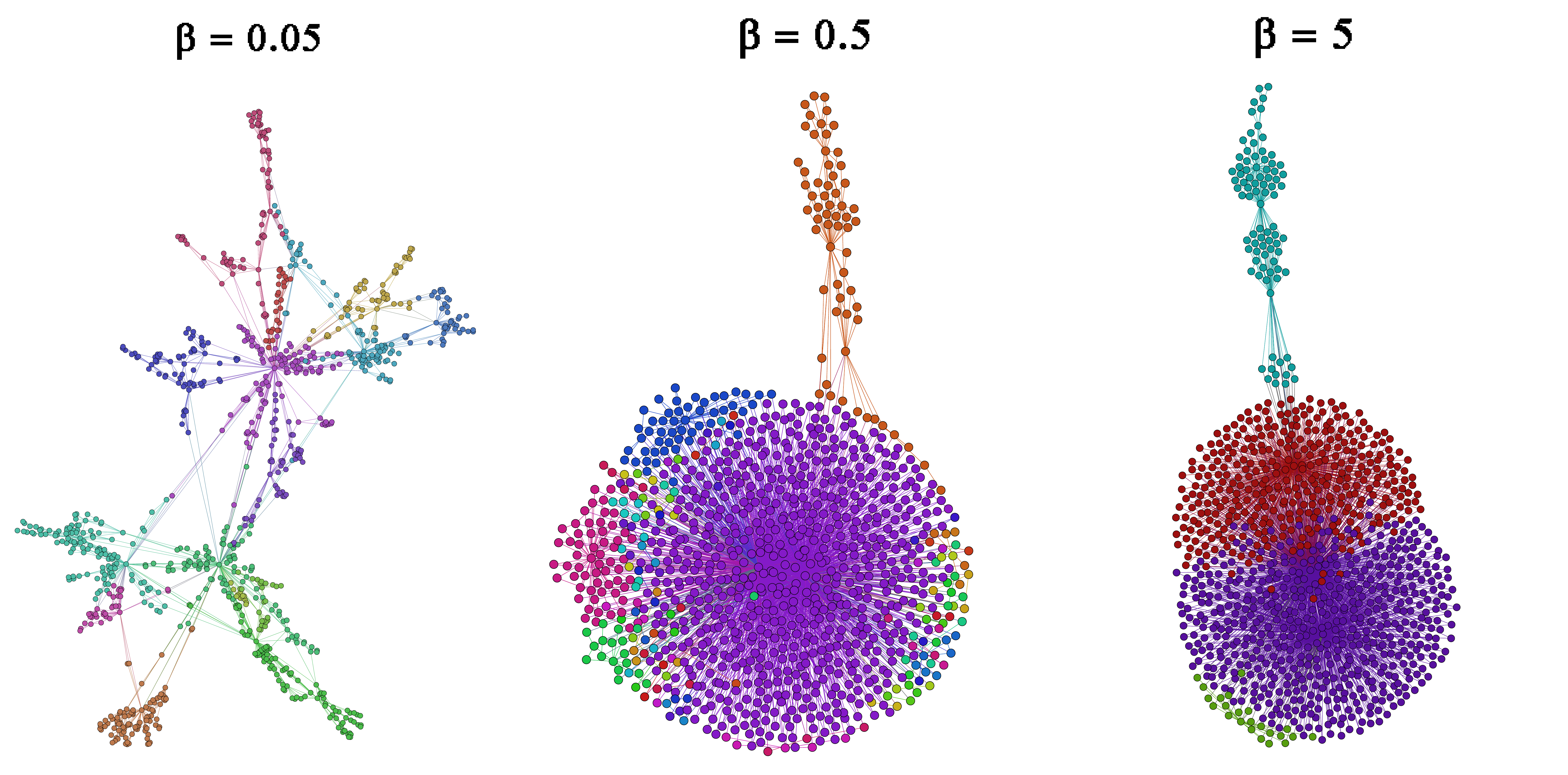}
	\caption{(Color online) Visualization of the  Bose-Einstein Network   (with  $g(\omega)$ given by Eq. $(\ref{go1})$   and $c=10$)  for $\beta=0.05,0.5,5$ and $N=1000$. For low value of $\beta$, i.e. 
	$\beta<\beta_c \simeq0.06$, the network is small-world, for large values of $\beta$, i.e. $\beta>\beta_c \simeq 0.06$, the network is condensed and develops a finite diameter. 
	The colour indicates the partition into communities found by running the Louvain algorithm \cite{Louvain}.
	}
	\label{visualization_m1}
\end{figure*}

\begin{figure*}	
\includegraphics[width=1.5\columnwidth]{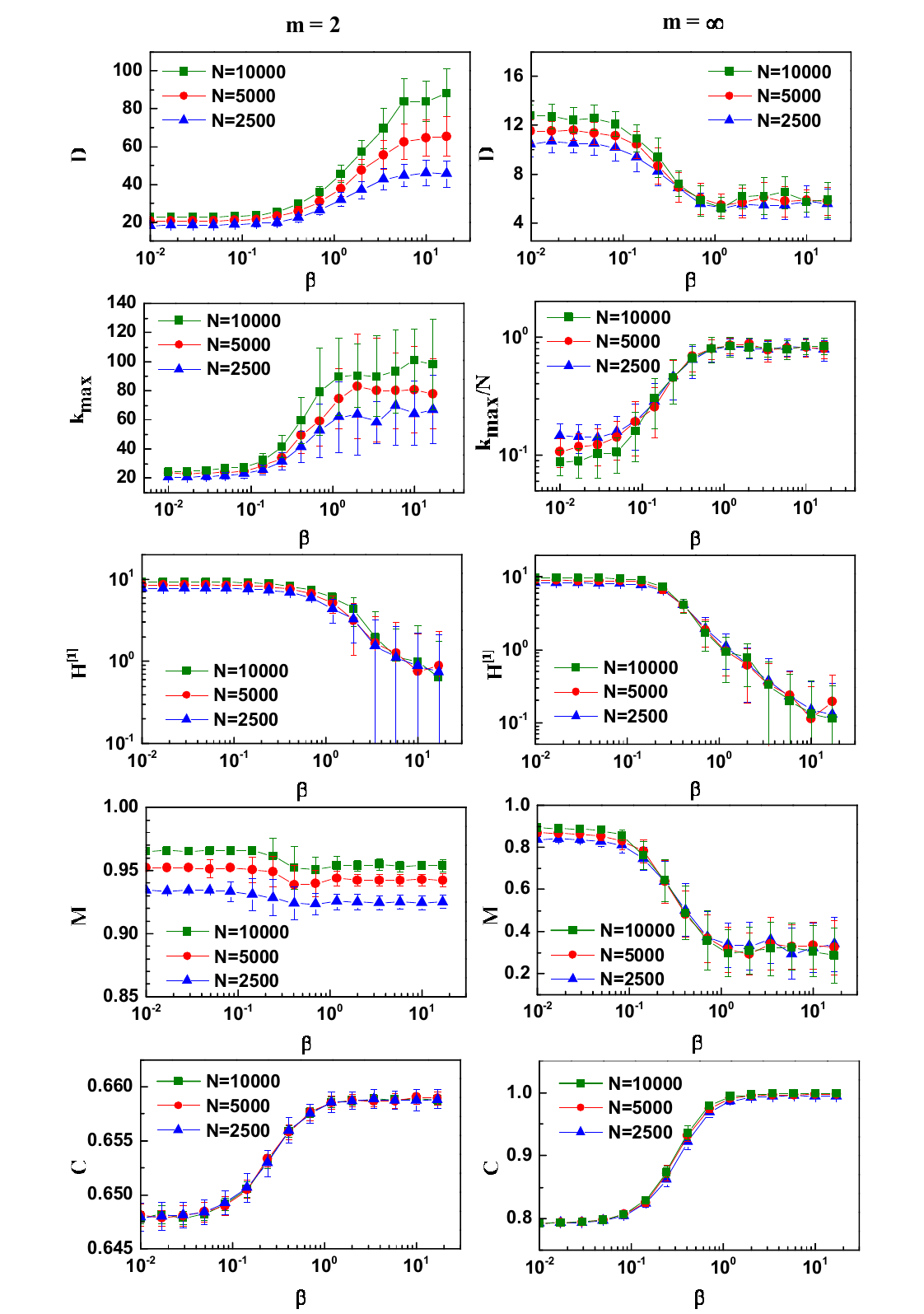}
\caption{(Color online) The maximal distance $D$ from the initial triangle, the maximal degree $k_{max}$,  the entropy rate  $H^{[1]}$, the modularity $M$ calculated using the Louvain algorithm \cite{Louvain}, and the average clustering coefficient 
$C$ are plotted as a function of the inverse temperature $\beta$ for the Fermi-Dirac Network ($m=2$) and for the Bose-Einstein Network ($m=\infty$). The networks have nodes with energies following a 
Poisson distribution $g(\omega)$ with average $c=10$. The data are reported for networks of size $N=10,000$  (averaged $30$ times), $N=5000$ 
(averaged $60$ times) and $N=2500$ (averaged $90$ times). The predicted phase transition for  the Fermi-Dirac Network  is at $\beta_c \simeq 0.14$, for the Bose-Einstein Network it is at $\beta_c \simeq 0.06$.  }
\label{transition_p0}
\end{figure*}

\begin{figure*}
\includegraphics[width=1.5\columnwidth]{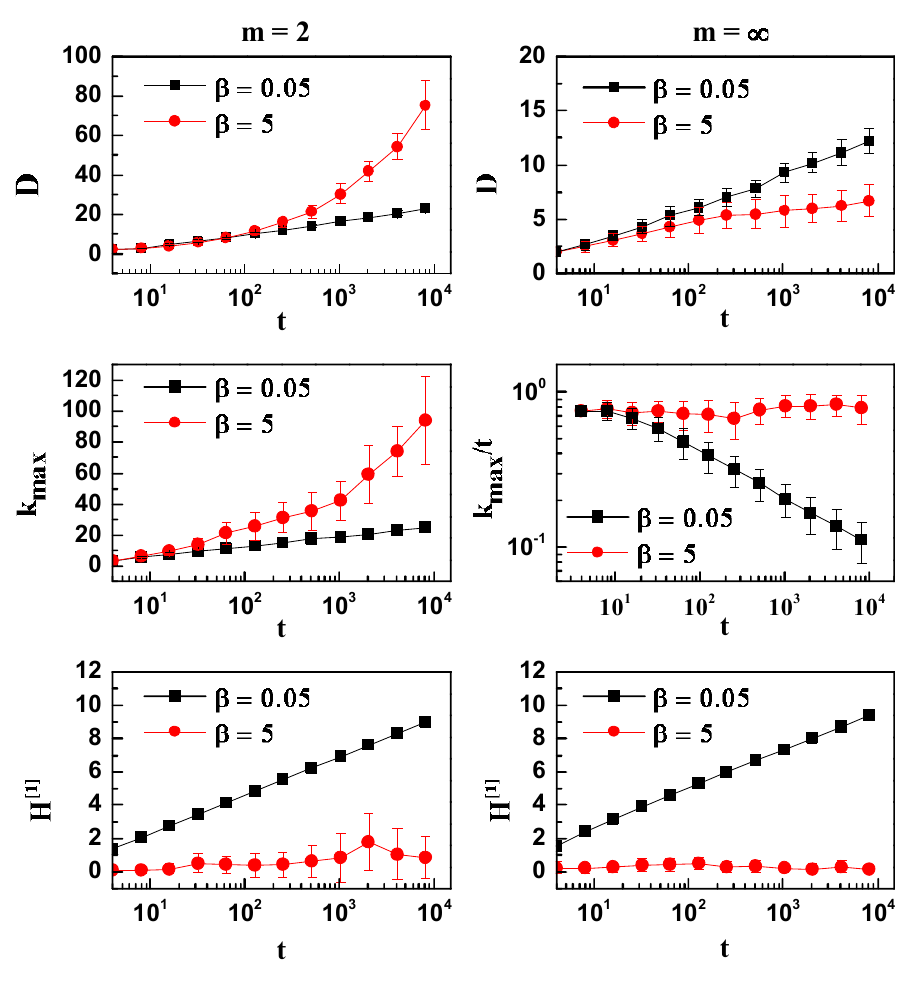}
	\caption{(Color online) The maximal distance $D$ from the initial triangle, the maximal degree $k_{max}$, and the entropy rate  $H^{[1]}$ are plotted as a function of time $t$ for the Fermi-Dirac Network ($m=2$) and for the 
	Bose-Einstein Network ($m=\infty$). The inverse temperatures are  $\beta=0.05$ and $\beta=5$ respectively, below and above the phase transitions.	
	The networks have nodes with energies following a Poisson distribution $g(\omega)$ with average $c=10$. The data are averaged $20$ times. }
	\label{transition_time_p0}
\end{figure*}

\begin{figure}
	\includegraphics[width=0.95\columnwidth]{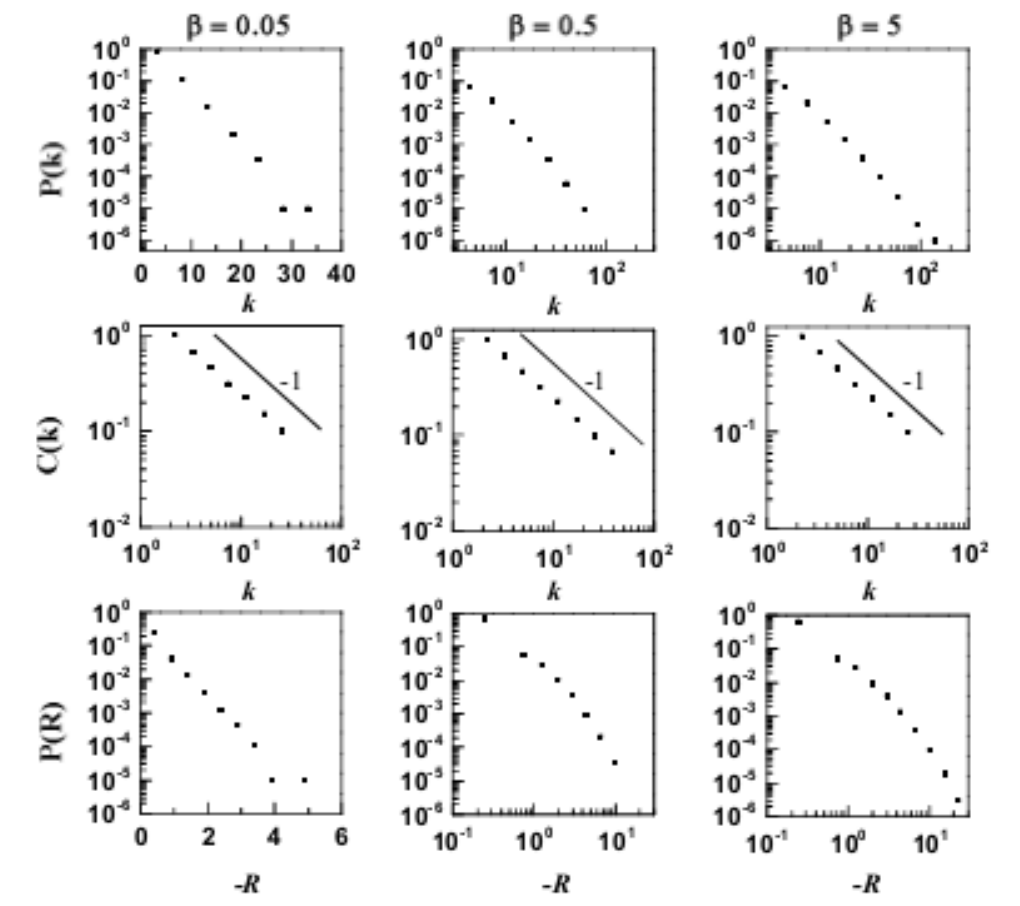}
	\caption{Structural and geometrical properties of the Fermi-Dirac Network as a function of the inverse temperature $\beta$ for single network realizations of size $N=10^5$. The degree distribution $P(k)$, the average clustering coefficient $C(k)$ of 
	nodes of degree $k$ and  the distribution of the  curvatures $P(R)$ are plotted for $\beta=0.05,0.5,5$. The networks have nodes with energies following a Poisson distribution $g(\omega)$ 
	with average $c=10$. For low values of $\beta$, i.e. $\beta<\beta_c\simeq 0.14$, 
	$P(k)$ and 
	$P(R)$ are exponential, while for large values of $\beta$, i.e. $\beta>\beta_c\simeq 0.14$ they become power-law. The average clustering coefficient
	$C(k)$ of nodes of degree $k$
	always goes like $C(k)\propto k^{-1}$.}	
	\label{figuresimm2}
\end{figure}

\begin{figure}
	\includegraphics[width=0.95\columnwidth]{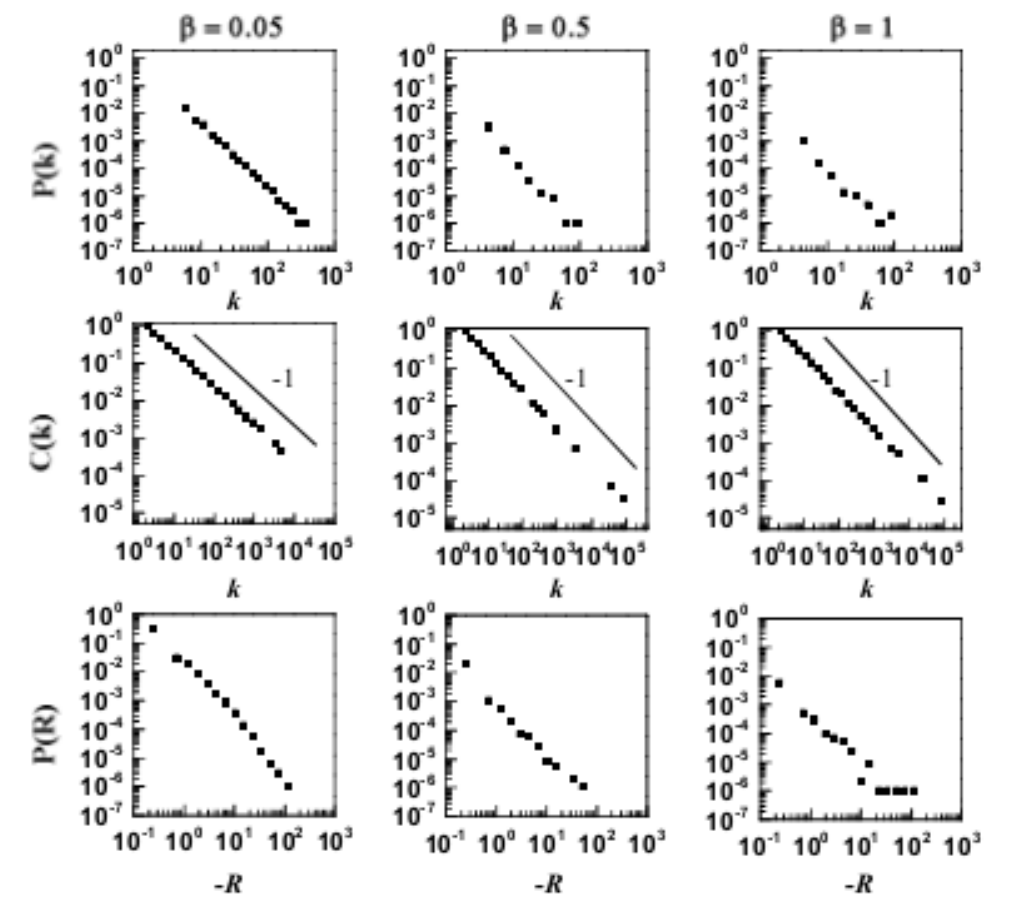}
	\caption{ Structural and geometrical properties of the Bose-Einstein Network as a function of the inverse temperature $\beta$ for single network realizations of size $N=10^5$. The degree distribution $P(k)$, the average clustering coefficient $C(k)$ of 
	nodes of degree $k$ and  the distribution of the  curvatures $P(R)$ are plotted for $\beta=0.05,0.5,1$. The networks have nodes with energies following a Poisson distribution $g(\omega)$ 
	with average $c=10$. For low values of $\beta$,  i.e. $\beta<\beta_c\simeq 0.06$,  
	$P(k)$ and 
	$P(R)$ are scale-free, while for large values of $\beta$, i.e. $\beta>\beta_c\simeq 0.06$, these distributions become dominated by outliers. 
	The average clustering coefficient 
	$C(k)$ 
	of nodes of degree $k$ 
	always goes like $C(k)\propto k^{-1}$. }	
	\label{figuresimm1}
\end{figure}

As a function of $\beta$ we observe a phase transition in both the Fermi-Dirac Network 
and the Bose-Einstein Network.
For $\beta>\beta_c$ the structure of the network and its geometry change drastically as can already be seen from the visualizations of the networks (Figure $\ref{visualization_m2}$ for the 
Fermi-Dirac Network and Figure $\ref{visualization_m1}$ for the Bose-Einstein Network).
The transition is characterized by a different scaling of the entropy rate $H^{[1]}$ below and above $\beta_c$. For $\beta<\beta_c$ $H^{[1]}$ increases with time as $H^{[1]}\simeq \ln(t)$, 
due to the linear scaling of $Z\propto t$, while, for $\beta>\beta_c$,  $H^{[1]}={\cal O}(1)$ and fluctuates widely during the network evolution.
Here we discuss in detail the consequences of this transition in the Fermi-Dirac Network 
and in the Bose-Einstein Network. 
In Figure $\ref{transition_p0}$ we show major geometrical and structural properties of the network as a function of the inverse temperature $\beta$ across the phase transitions. In particular we 
display the maximal shortest (hopping) distance from a given node of the  initial triangle $D$, the maximal degree $k_{max}$ of the network, the entropy rate $H^{[1]}$, the modularity $M$ \cite{Newman} calculated using the Louvain algorithm \cite{Louvain}  
and the average clustering coefficient $C$ across the phase transitions.
In Figure $\ref{transition_time_p0}$ we show major geometrical and structural properties of the network as a function of time for given values of the inverse temperature $\beta$ below and 
above the phase transition.
Finally in Figure $\ref{figuresimm2}$ and Figure $\ref{figuresimm1}$ we show the degree distribution $P(k)$,  the average clustering coefficient $C(k)$ of nodes of degree $k$,  and the 
distribution of the curvature $P(R)$, for the Fermi-Dirac and the Bose-Einstein Network above and below the phase transition.

For the  Fermi-Dirac Network, the most important indicator of the phase transition is $D$ which grows logarithmically with time for
$\beta<\beta_c$,  and as a power-law for $\beta>\beta_c$. Therefore the network is small-world for $\beta<\beta_c$ while it has finite Hausdorff dimension  for $\beta>\beta_c$. Moreover the maximum degree $k_{max}$ increases significantly below the transition for $\beta>\beta_c$.

Furthermore, for $\beta<\beta_c$,  the degree distribution $P(k)$ is exponential, and the distribution of the curvature 
$P(R)$ has a negative exponential tail,the average curvature is $\Avg{R}=1/N$ and its second moment $\Avg{R^2}$ is finite.\\
For $\beta>\beta_c$, instead, $P(k)$ follows a power-law, and
$P(R)$ has a negative power-law tail. In this case, the average 
curvature is $\Avg{R}=1/N$, but its second moment $\Avg{R^2}$ diverges. For every value of $\beta$,  the network has high modularity $M$ and a hierarchical structure \cite{Ravasz} 
with an average clustering coefficient $C(k)$ of nodes of degree $k$ decaying as $C(k)\simeq k^{-\alpha}$ and $\alpha=1$.

For the  Bose-Einstein Network, the most important indicator of the phase transition is the maximum degree $k_{max}$ which scales sub-linearly with time for
$\beta<\beta_c$ and linearly for $\beta>\beta_c$, i.e.  in this case the most connected node is linked to a finite fraction of all the nodes. 
Moreover, for $\beta<\beta_c$, $D$ 
increases logarithmically with the network size, i.e. the network is  small-world, while for $\beta>\beta$ it decreases significantly.

Furthermore, for $\beta<\beta_c$, the degree distribution $P(k)$ is scale-free,
the network has high modularity $M$,
and  the distribution of the curvature $P(R)$ has a negative power-law tail.
For $\beta>\beta_c$, instead,
$P(k)$ is dominated by outlier hubs, 
the network has low modularity $M$ 
and $P(R)$ has a negative tail dominated by outlier nodes.\\
In both cases, the average curvature is $\Avg{R}=1/N$, and its second moment $\Avg{R^2}$ diverges.
The network has a hierarchical structure \cite{Ravasz} with an average clustering coefficient $C(k)$ of nodes of degree $k$ decaying as $C(k)\simeq k^{-\alpha}$ and $\alpha=1$.




\section{Quantum network states}
\label{cao}
\subsection{The Hilbert space}
Using a similar approach used already in \cite{graphity_rg,graphity,graphity_c}, here we define quantum network states.
In \cite{graphity_rg,graphity,graphity_c}  an Hilbert space is associated to each node and each possible link of a network of $N$ nodes.
Here we associate to each node and Hilbert space ${\cal H}_{node}$ and to each link we associate two  Hilbert spaces ${\cal H}_{link}$ and $\tilde{{\cal H}}_{link}$.
The total Hilbert space ${\cal H}_{N}$of a network of $N$ nodes is given by 
\bea
{\cal H}_N=\bigotimes^N {\cal H}_{node} \bigotimes^{N(N-1)/2}{\cal H}_{link} \bigotimes^{N(N-1)/2} \tilde{{\cal H}}_{link}.
\eea

A single realization of a growing geometric  network of size $N$ in which the nodes are labelled by the time they have been added to the network can be mapped to a quantum state by mapping the nodes, the links and the triangles of the network to quantum states as described in the following.

\subsection{Nodes quantum states}

To every node $i$ we associate a Hilbert space ${\cal H}_{node}$ which  the Hilbert space of a fermionic oscillator with energy   $\omega_i$.
Therefore, to every node $i$ of the network we associate a  {\it node quantum state} that can be decomposed in the basis
\bea
\{\ket{o_i,\omega_i}\},
\eea
with $o_i=0,1$.
The state
\bea
\ket{o_i=1,\omega}
\eea
is said to contain a particle of energy $\omega$ and  
can be mapped to the presence of  the node $i$  with energy $\omega=\omega_i$ in the network.
The state
\bea
\ket{o_i=0,\omega}
\eea 
is an empty state and 
can be mapped to the absence of  the node $i$ in the network. In this case the value of $\omega$ is irrelevant to characterize the state.

\subsection{Link quantum states}
To every possible link $(i,j)$ of a network we associate an Hilbert space ${\cal H}_{link}$.The Hilbert space ${\cal H}_{link}$ is chosen to be that of a fermonic oscillator. To every pair of nodes $(i,j)$ in the network we associate a  {\it link quantum state} that can be decomposed in the basis
\bea
\{\ket{a_{ij}}\},
\eea
where $a_{ij}=0,1$.
The state 
\bea
\ket{a_{ij}=1}
\eea
is said to contain a particle and 
is mapped to a link $(i,j)$ in the network.
The state 
\bea
\ket{a_{ij}=0}
\eea
is an empty state and 
is mapped to the absence of a link in the network.
\subsection{Incident triangles quantum state}
To every possible link $(i,j)$ of a network we associate an Hilbert space ${\widetilde{\cal H}}_{link}$. For the Fermi-Dirac network state we will assume that this Hilbert space is the one associated to a fermionic oscillator. For the Bose-Einstein network state we will assume that this Hilbert space is instead associated with a bosonic oscillator.  
Therefore to each possible link of the network we associate a   {\it incident triangles quantum state}  that  can be decomposed in the basis
\bea
\{\ket{n_{ij}}\},
\eea
with $n_{ij}=0,1$ for the Fermi-Dirac Network and $n_{ij}=0,1,2,\ldots$ for the Bose-Einstein Network. 
The quantum number  $n_{ij}$  of links for which $a_{ij}=1$, is mapped to the number of triangles exceeding one incident to every existing link $(i,j)$.
Therefore if $m=2$ we can only have $n_{ij}=0,1$ while if $m=\infty$ we can have any integer value of the occupation number $n_{ij}=0,1,2\ldots$.

\subsection{Quantum network states}
At each time $t$ the quantum network state can be decomposed into a basis 
\bea
\hspace*{-5mm}\ket{\{o_i,\omega_i,a_{ij},n_{ij}\}}=\prod_i \ket{o_i, \omega_i}\prod_{i<j} \ket{a_{ij}}\prod_{i<j}\ket{n_{ij}}\ .
\eea
We consider the following operators that act on the nodes, links and incident triangles quantum states.

\subsection{Creation-Annihilation operators}
The operators $b^{\dag}_{j}(\omega),b_{i}(\omega)$    are creation-annihilation of  node quantum  states and have  anti-commutation relations
\bea
\{b_{i}(\omega),b^{\dag}_{j}(\omega^{\prime})\}&=&\delta(i,j)\delta(\omega,\omega^{\prime}),\nonumber \\
\{b_{i}(\omega),b_{j}(\omega^{\prime})\}&=&0,\nonumber \\
\{b^{\dag}_{i}(\omega),b^{\dag}_{j}(\omega^{\prime})\}&=&0,
\eea 
where $\delta(x,y)$ is the Kronecker delta, $\delta(x,y)=1$ for $x=y$ and $\delta(x,y)=0$ otherwise.
They act on the node states $\{\ket{o_i,\omega_i}\}$ as
\bea
b_{i}(\omega_i)\ket{o_{i}=0,\omega_i}&=&0,\nonumber \\
b_{i}(\omega_i)\ket{o_{i}=1, \omega_i}&=&\ket{o_{i}=0,\omega_i},\nonumber \\
b^{\dag}_{i}(\omega_i)\ket{o_{i}=0,\omega_i}&=&\ket{o_{i}=1,\omega_i},\nonumber \\
b^{\dag}_{i}(\omega_i)\ket{o_{i}=1,\omega_i}&=&0.
\eea

The operators $c^{\dag}_{ij},c_{ij}$  are creation-annihilation operator of  link quantum  states and   have anti-commutation relations
\bea
\{c_{ij},c^{\dag}_{rs}\}&=&\delta[{(i,j),(r,s)}],\nonumber \\
\{c_{ij},c_{rs}\}&=&0,\nonumber \\
\{c^{\dag}_{ij},c^{\dag}_{rs}\}&=&0.
\eea 
where $\delta[(i,j),(r,s)]=1$ if $i=r$ and $j=s$ or if $i=s$ and $j=r$ and $\delta[(i,j),(r,s)]=0$ otherwise.
They act on the link states $\ket{a_{ij}}$ as
\bea
c_{ij}\ket{a_{ij}=0}&=&0,\nonumber \\
c_{ij}\ket{a_{ij}=1}&=&\ket{a_{ij}=0},\nonumber \\
c^{\dag}_{ij}\ket{a_{ij}=0}&=&\ket{a_{ij}=1},\nonumber \\
c^{\dag}_{ij}\ket{a_{ij}=1}&=&0.
\eea

Finally we define two classes of creation annihilation operators  acting respectively on the incident triangles quantum states of Fermi-Dirac and Bose Einstein quantum network states.
The creating and annihilation operators $d^{\dag}_{ij}$ and $d_{ij}$ acting on incident triangle quantum states of Fermi-Dirac network states are anti-commuting,i.e. 
\bea
\{d_{ij},d^{\dag}_{rs}\}&=&\delta[{(i,j),(r,s)}],\nonumber \\
\{d_{ij},d_{rs}\}&=&0,\nonumber \\
\{d^{\dag}_{ij},d^{\dag}_{rs}\}&=&0.
\eea 
When these operators act  on the incident triangles quantum states, they can only generate occupation numbers $n_{ij}=0,1$.
Their action on the basis $\{\ket{n_{ij}=0},\ket{n_{in}=1}\}$ is given by 
\bea
d_{ij}\ket{n_{ij}=0}&=&0,\nonumber \\
d_{ij}\ket{n_{ij}=1}&=&\ket{n_{ij}=0},\nonumber \\
d^{\dag}_{ij}\ket{n_{ij}=0}&=&\ket{n_{ij}=1},\nonumber \\
d^{\dag}_{ij}\ket{n_{ij}=1}&=&0.
\eea
The creation annihilation operators $\tilde{d}^{\dag}_{ij}$ and $\tilde{d}_{ij}$ that are acting on the incident triangles quantum states of the Bose-Einstein quantum network states  have the commutation relations 
\bea
\left[ \tilde{d}_{ij},\tilde{d}^{\dag}_{rs}\right]&=&\delta[{(i,j),(r,s)}],\nonumber \\
\left[ \tilde{d}_{ij},\tilde{d}_{rs}\right]&=&0,\nonumber \\
 \left[\tilde{d}^{\dag}_{ij},\tilde{d}^{\dag}_{rs} \right]&= & 0.
\eea
When these operators act  on the incident triangles quantum states, they can  generate arbitrary   occupation numbers $n_{ij}=0,1,2, \ldots$
Their action on the basis $\{\ket{n_{ij}=n}\}$ is given by 
\bea
\tilde{d}_{ij}\ket{n_{ij}=n}&=&\sqrt{n}\ket{n_{ij}=n-1},\nonumber \\
\tilde{d}^{\dag}_{ij}\ket{n_{ij}=n}&=&\sqrt{n+1}\ket{n_{ij}=n+1}.\nonumber \\
\eea

\section{Evolution of quantum network states}
In this section we define a non equilibrium  Markovian evolution of the quantum network states. The possibility of a Markovian evolution of quantum network states is not entirely new in the literature, as it has been for example proposed in \cite{graphity_rg}.
Therefore we will  define a  quantum network state $\ket{\psi_N(t)}$ and its evolution with time.
The quantum network state at every time step can be decomposed into the base   
\bea
\ket{\psi_N(t)} =\sum_{\{o_i,\omega_i,a_{ij},n_{ij}\}} C_{\{o_i,\omega_i,a_{ij},n_{ij}\}}\ket{ \{o_i,\omega_i,a_{ij},n_{ij}\}}.
\nonumber\eea
We start from an initial condition at $t=1$ given by 
\bea
\hspace*{-4mm}\ket{\psi_N(1)} =\frac{1}{\sqrt{{\cal Z}(1)}}\sum_{\omega_1,\omega_2,\omega_3} \left[\prod_{i=1,2,3}\rho(\omega_i)b^{\dag}_i(\omega_i)\right]c^{\dag}_{12}c^{\dag}_{23}c^{\dag}_{13}\ket{0},\nonumber
\eea
where ${\cal Z}(1)$ is fixed by the normalization condition $\langle \psi_N(1)|\psi_N(1)\rangle =1$.
The quantum evolution of the network state is given by a Markov process whose transition rate is determined by the unitary operator $U$
\bea
&&\ket{\psi_N(t)}=U_t\ket{\psi_N(t-1)},
\label{psiNt}
\eea
where $U_t$ is defined by 
\bea
U_t&&=\sqrt{\frac{{\cal Z}(t-1)}{{\cal Z}(t)}}\sum_{\omega_{t+2}}\sum_{i,j|i<j}\sqrt{g(\omega_{t+2})}e^{-\beta \epsilon_{ij}/2}\nonumber \\
&&\times b^{\dag}_{t+2}(\omega_{t+2})c^{\dag}_{(t+2)i}c^{\dag}_{(t+2)j}h^{\dag}_{ij}c^{\dag}_{ij}c_{ij}.
\label{psiNt2}
\eea
Here as in Eq. (\ref{unob}) $\epsilon_{ij}=f(\omega_i,\omega_j)$ and ${\cal Z}(t)$ is fixed by the normalization condition $\Avg{\psi_N(t)|\psi_N(t)}=1$.
The quantum operators $h^{\dag}_{ij}$, can take two different values, defining in this way the Fermi-Dirac Quantum Network state and the Bose-Einstein Quantum Network state, i.e.
\bea
h^{\dag}_{ij}=\left\{\begin{array}{ll} d^{\dag}_{ij} & \mbox{Fermi-Dirac Quantum Network state}\nonumber \\
\tilde{d}^{\dag}_{ij} & \mbox{Bose-Einstein Quantum Network state}\end{array}\right. .
\eea
In the following section we will consider in detail the Fermi-Dirac and the Bose-Einstein Quantum Network states. 
\section{Fermi-Dirac Quantum Network  Evolution}

\subsection{Path integral}
The evolution of the Fermi-Dirac Quantum Network state is given by 
\bea
&&\ket{\psi_N(t)}=U_t\ket{\psi_N(t-1)}\nonumber \\
&&=\sqrt{\frac{{\cal Z}(t-1)}{{\cal Z}(t)}}\sum_{\omega_{t+2}}\sum_{i,j|i<j}\sqrt{g(\omega_{t+2})}e^{-\beta \epsilon_{ij}/2}\nonumber \\
&&\times b^{\dag}_{t+2}(\omega_{t+2})c^{\dag}_{(t+2)i}c^{\dag}_{(t+2)j}d^{\dag}_{ij}c^{\dag}_{ij}c_{ij}\ket{\psi_N(t-1)}
\eea
where  $\epsilon_{ij}=f(\omega_i,\omega_j)$ and ${\cal Z}={\cal Z}(t)$ is fixed by the normalization condition $\langle{\psi_N(t)}|{\psi_N(t)}\rangle=1$. 

Using the definitions of the creation and annihilation operators defined in Sec. \ref{cao}, the normalization constant ${\cal Z}(t)$ of the Fermi-Dirac Quantum Network state is fixed by the path integral
\bea
{\cal Z}(t)&=&\sum_{\{\omega(t')\}}\sum_{\{\ell(t')\}} W( \{\omega({t'}),\ell(t')\}_{t'\leq t})
\label{ZF}
\eea
for $t\geq 2$.
Here $\{\ell_{t'}\}_{t'=1,\ldots, t}$ is a sequence of links $\ell(t')=(i_{t'},j_{t'})$ and $\{\omega({t'})\}_{t'=1,\ldots, t}$ is a sequence of energies of the nodes that describes a single history over which the 
path integral is calculated. In the path integral in Eq. $(\ref{ZF})$  each path  $\{\omega({t'}),\ell(t')\}_{t'=1,\ldots, t}$ is assigned a weight
 \bea
 W( \{\omega({t'}),\ell(t')\}_{t'\leq t})&=&\prod_{i=1}^{t+2}g(\omega_i)\prod_{t'\leq t}a_{\ell(t')}(t')\nonumber \\
&&\hspace*{-30mm}\times(1-n_{\ell(t')}(t'))e^{-\beta\sum_{i<j}\epsilon_{ij} n_{ij}(t)},
\label{Pp}
 \eea
where the terms  $a_{ij}(t)$ and $n_{ij}(t)$ that appear in Eq. $(\ref{Pp})$ can be expressed in terms of the history $\{\ell(t')\}_{t'\leq t}$ as
\bea\label{FDaandn}
a_{ij}(t)&=&\sum_{t'= 2}^{t-1}\left(\delta[(i,j),(t'+2,i_{t'})]+\delta[(i,j),(t'+2,j_{t'})]\right)\nonumber \\
&&+\delta[(i,j),(1,2)]+\delta[(i,j),(1,3)]+\delta[(i,j),(2,3)],\nonumber \\
 n_{ij}(t)&=&\sum_{t'=2}^{t-1}\delta[\ell(t'),(i,j)].
 \eea
 Therefore ${\cal Z}(t)$ can be interpreted as a partition function of a statistical mechanics problem in which each path up to time $t$ has probability 
 \bea
 P( \{\omega({t'}),\ell(t')\}_{t'\leq t})=\frac{W( \{\omega({t'}),\ell(t')\}_{t'\leq t})}{{\cal Z}(t)}.
 \label{Ppb}
 \eea

Each of the  paths $\{\omega({t'}), \ell(t')\}_{t'\leq t}$ can be mapped to a geometrical network evolution 
  with $m=2$. 
  In this mapping $\omega(t)$ indicates the energy of the node added to the network at time $t$, $\ell(t)=(i_t,j_t)$ indicates the link to which we attach a new triangle at time $t$, $a_{ij}(t)$ indicates the adjacency 
  matrix of the network, $n_{ij}(t)$ indicates the additional number of triangles incident to an existing  link $(i,j)$.
  The probability of each geometrical network evolution $\{\omega({t'}),\ell(t')\}_{t'=1,\ldots, t}$ described in Sec. \ref{gmb} is the same as the weight that the corresponding history has 
  in Eq. $(\ref{Pp})$.
  
Starting from Eq. $(\ref{Ppb})$ we can calculate the conditional probability  that at time $t$ we add a link $\ell(t)=(i,j)$ given the present state of the network evolution,  
$\Pi^{[F]}_{(i,j)}=P\left(\ell(t)=(i,j)|\{\omega({t'}),\ell(t')\}_{t'<t}\right)$. A straightforward calculation shows that 
\bea
\Pi^{[F]}_{(i,j)}(t)=\frac{e^{-\beta\epsilon_{ij}}a_{ij}(t)(1-n_{ij}(t))}{Z_F},
\label{probF}
\eea
where 
\bea
Z_F=\sum_{i<j}e^{-\beta\epsilon_{ij}}a_{ij}(t)(1-n_{ij}(t)).
\label{Zb}
\eea
Given that $1-n_{ij}=1$ has the graphical network interpretation $\xi_{ij}=1$, i.e. indicates that the link $(i,j)$ is not saturated, while $1-n_{ij}=0$ indicates that the link is saturated, i.e. $\xi_{ij}=0$, the expression in Eq. $(\ref{probF})$ is the same as $\Pi_{(i, j)}^{[1]}$ defined in Eq. $(\ref{prob})$. 
It follows that studying the geometrical network evolution for $m=2$ determines the properties of the Fermi-Dirac quantum network state.
For this reason we call the growing geometrical network with $m=2$ the Fermi-Dirac Network.

\subsection{Fermi-Dirac Statistics}
The average of the quantum number $n_{ij}$ over all the links of the Fermi-Dirac Network follows the Fermi-Dirac distribution. Since $n_{ij}=0,1$, equivalently, we can say that the probability that a link with energy 
$\epsilon$ is saturated follows the Fermi-Dirac statistics.
To derive this result, let us consider the master equation \cite{Doro_book} for the number $N_F^{t}(n|\omega,\omega')$ of  links $(i,j)$   (with $i>j$, $\omega_i=\omega $ and $\omega_j=\omega^{\prime}$), that have  $n_{ij}=n=0,1$ at time $t$.
 Since at each time we choose a  link $(i,j)$ with probability $\Pi^{[F]}_{(i,j)}(t)$, only if it is unsaturated (i.e. $n_{ij}=0$), we add one triangle to the link (i.e.  $n_{ij}=0\to n_{ij}=1$) 
 and we add other two unsaturated links, the master equation reads
\bea
{ N_F^{t+1}(n=1|\omega,\omega')}&=&\frac{e^{-\beta \epsilon}}{Z_F}N_F^t({n=0}|\omega,\omega')\nonumber \\
&&+N_F^{t}(n=1|\omega,\omega')\nonumber \\
{N_F^{t+1}(n=0|\omega,\omega')}&=&-\frac{e^{-\beta \epsilon}}{Z_F}N_F^t(n=0|\omega,\omega')\nonumber \\
&&\hspace*{-10mm}+2\rho_F(\omega,\omega')+N_F^{t}(n=0|\omega,\omega'),
\label{NFn}
\eea
where $\epsilon=f(\omega,\omega^{\prime})$ and $\rho_F(\omega,\omega')$ is the probability that  a new link $(i,j)$ of the network with $i>j$ links two nodes with energy $\omega_i=\omega$ and $\omega_j=\omega^{\prime}$.
In order to solve this master equation we assume that the normalization constant $Z_F\propto t$ and we put
\bea
e^{-\beta \mu_F}&=&\lim_{t\to \infty}\frac{Z_F}{t}.
\label{selfF0}
\eea
This is a self-consistent assumption that must be verified by the solution of Eqs. $(\ref{NFn})$.
Moreover we also assume that at large times $N_F^t(n|\omega,\omega')\simeq 2tP_F(n|\omega,\omega')$. In fact the number of links in the network is $2t+1\simeq 2t$ for $t\gg 1$.
Here $P_F(n|\omega,\omega')$ indicates the asymptotic probability that a random link $(i,j)$ with $i>j$ and $\omega_i=\omega$, $\omega_j=\omega^{\prime}$ has $n_{ij}=n$. With these assumptions, 
we can solve Eqs. $(\ref{NFn})$ finding
\bea
\hspace*{-10mm}P_F(n=0|\omega,\omega')&=&\rho_F(\omega,\omega')\frac{e^{\beta(\epsilon-\mu_F)}}{e^{\beta(\epsilon-\mu_F)}+1}\nonumber \\
&=&\rho_F(\omega,\omega')[1-n_F(\epsilon)]\nonumber \\
\hspace*{-10mm}P_F(n=1|\omega,\omega')&=&\rho_F(\omega,\omega')\frac{1}{e^{\beta(\epsilon-\mu_F)}+1}\nonumber \\
&=&\rho_F(\omega,\omega')n_F(\epsilon)
\label{NsF}
\eea
where $\epsilon=f(\omega,\omega^{\prime})$ and $n_F(\epsilon)$ is the Fermi-Dirac occupation number \bea
n_F(\epsilon)=\frac{1}{e^{\beta(\epsilon-\mu_F)}+1}.
\eea
Considering all the links with energy 
$\epsilon$, we have 
\bea
\avg{n|\epsilon}=\rho_F(\epsilon)n_F(\epsilon),
\label{nFno}
\eea
where
\bea
\rho_F(\epsilon)=\sum_{\omega,\omega'}\delta[\epsilon,f(\omega,\omega')]\rho_F(\omega,\omega')
\eea
and where 
\bea
\hspace*{-10mm}\avg{n|\epsilon}=\sum_{\omega,\omega^{\prime}}\delta[\epsilon,f(\omega,\omega')]\rho_F(\omega,\omega^{\prime})\sum_{n=0,1}n P_F(n|\omega,\omega^{\prime}).
\eea
Therefore, in the Fermi-Dirac Network the average of the incident triangles quantum number over links of energy $\epsilon$ follows the Fermi-Dirac distribution.

To complete the solution  it is necessary to find the correct expression for $\rho_F(\omega,\omega')$.
Since by definition  $\omega$ is the energy of the new node attached to the network at time $t$, and since this energy is drawn randomly from a distribution $g(\omega)$, 
we have that the probability $\rho_F(\omega,\omega^{\prime})$ can be factorized,
\bea
\rho_F(\omega,\omega')=g(\omega)\tilde{g}_F(\omega^{\prime}),
\eea
where $\tilde{g}_F(\omega^{\prime})$ is the probability that a new triangle is attached to a link having at its end a  node of  energy $\omega^{\prime}$ and is thus normalized.
Therefore we can write a recursive equation for $\tilde{g}_F(\omega^{\prime})$. Since we attach new triangles to random unsaturated links with energy $\epsilon_{ij}=\epsilon=f(\omega, \omega')$ 
with probability $e^{-\beta(\epsilon-\mu_F)}$, it follows that the recursive equation for $\tilde{g}_F(\omega)$ reads
\bea
\tilde{g}_F(\omega)&=&\sum_{\omega^{\prime}}e^{-\beta[f(\omega,\omega')-\mu_F]}\left[P_F(n=0|\omega,\omega')\right. \nonumber \\
&&\left.+P_F(n=0|\omega',\omega)\right]\nonumber \\
&=&\sum_{\omega^{\prime}}\left[\rho_F(\omega,\omega^{\prime})+\rho_F(\omega^{\prime},\omega)\right]n_F[f(\omega,\omega')],
\label{rFu}
\eea
where in the last equation  we have used the expression for $P_F(n=0|\omega,\omega')$ given by Eq. (\ref{NsF}).
Eq. $(\ref{rFu})$ can be formulated as the eigenvalue problem
\bea
\tilde{g}_F(\omega)&=&\sum_{\omega^{\prime}}A_F(\omega,\omega^{\prime})\tilde{g}(\omega^{\prime}),
\eea
where 
\bea
A_F(\omega,\omega^{\prime})&=&g(\omega)n_F[f(\omega,\omega')]\nonumber \\
&&\hspace*{-5mm}\times \left\{1-\sum_{\omega^{\prime\prime}} g(\omega^{\prime\prime})n_F[f(\omega,\omega^{\prime\prime})]\right\}^{-1}.
\eea
Since we require that $\tilde{g}_F(\omega)$ is a probability, i.e. it is non-negative and normalized, the solution of the eigenvalue problem is given by the  Perron-Frobenious eigenvector $\tilde{g}_F(\omega)$ of the matrix $A_F(\omega,\omega^{\prime})$
satisfying 
\bea
\sum_{\omega}\tilde{g}_F(\omega)=1.
\eea
Finally the chemical potential $\mu_F$ is fixed by the self-consistent condition in Eq. $(\ref{selfF0})$ that can be rewritten as 
\bea
\sum_{\epsilon}\rho_F(\epsilon)n_F(\epsilon)=\frac{1}{2}
\label{selfF}
\eea
which is the same equation as the one fixing the chemical potential in a Fermi gas \cite{Kardar} with density of states $\rho_F(\epsilon)$, inverse temperature $\beta$ and specific volume $v=2$.
If the   self-consistent equation given by Eq. $(\ref{selfF0})$ has a solution, and $Z_F\propto t$,  the master equation asymptotically in time has a stationary solution given by Eqs. (\ref{NsF}). This implies that the  average of the quantum numbers $n_{ij}$ over links of energy $\epsilon$, i.e.  $\avg{n|\epsilon}$,  follows the Fermi-Dirac distribution.

\subsection{Structural properties of the Fermi-Dirac model}
Let us here characterize some of the important structural properties of the Fermi-Dirac network model.
First of all, let us consider the degree distribution $P(k)$.
In order to find $P(k)$ we first write the master equation  for the number $N_F^t(k|\omega)$ of nodes that at time $t$ have degree $k$ given that they have energy $\omega_i=\omega$. 
For simplicity in this paragraph we consider the linear relation Eq. (\ref{linearenergy}) between link and node energies.

The master equation \cite{Doro_book} for $N_F^t(k|\omega)$ reads
\bea
{N_F^{t+1}(k|\omega)}&=&\frac{e^{-\beta (\omega-\tilde{\mu}_F)}}{t}N_F^t(k-1|\omega)[1-\delta(k,2)]\nonumber \\
&&-\frac{e^{-\beta (\omega-\tilde{\mu}_F)}}{t}N_F^t(k|\omega)+g(\omega)\delta({k,2})\nonumber \\
&&+N_F^{t}(k | \omega)
\label{NFk}
\eea
where we have  assumed  that asymptotically in time we can define the chemical potential $\tilde{\mu}_F$ given by 
\bea
e^{\beta \tilde{\mu}_F}=e^{\beta \mu_F}\lim_{t \to \infty} \Avg{\frac{\sum_{ij}e^{-\beta \omega_j}(1-n_{ij})a_{ij}\delta(k_i,k)}{\sum_{i} \delta(k_i,k)}}.\nonumber
\eea
By assuming in the large network limit $t\gg1$ that $N_F(k|\omega)\simeq tP(k|\omega)$, solving Eq. ($\ref{NFk}$) we get, 
\bea
P(k|\omega)=g(\omega)\frac{e^{\beta(\omega-\tilde{\mu}_F)}}{\left[e^{\beta (\omega-\tilde{\mu}_F)}+1\right]^{k-1}}
\label{PkFc}
\eea
for $k\geq 2$.
Therefore, summing over all the values of the energy of the nodes $\omega$ we get the full degree distribution $P(k)$
\bea
P(k)=\sum_{\omega} g(\omega)\frac{e^{\beta(\omega-\tilde{\mu}_F)}}{\left[e^{\beta (\omega-\tilde{\mu}_F)}+1\right]^{k-1}}
\label{PkF}
\eea
for $k\geq 2$.

The curvature $R_i$ of a node $i$ is given by 
\bea
R_i=\frac{4-k_i}{6}.
\eea
So the distribution of the curvature $P(R)$ is given by 
\bea
P(R)=\sum_{\omega} g(\omega)\frac{e^{\beta(\omega-\tilde{\mu}_F)}}{\left[e^{\beta (\omega-\tilde{\mu}_F)}+1\right]^{3(1-2R)}}
\eea
where $R\leq \frac{1}{3}$. Therefore the distribution of the curvature is decaying exponentially for  negative values of the  curvature.Moreover the average curvature is $\Avg{R}=1/N$ and the fluctuations around this average are bounded,i.e. $\Avg{R^2}<\infty$.

\subsection{Comparison with numerical simulations}
Here we numerically simulate a Fermi-Dirac Network Evolution in which the energies of the nodes are non-negative integers with $g(\omega)$ given by a Poisson distribution with average $c$
as in Eq. (\ref{Poissondistribution}) and link and node energies related linearly Eq. (\ref{linearenergy}).

\begin{figure*}
	\includegraphics[width=1.65\columnwidth]{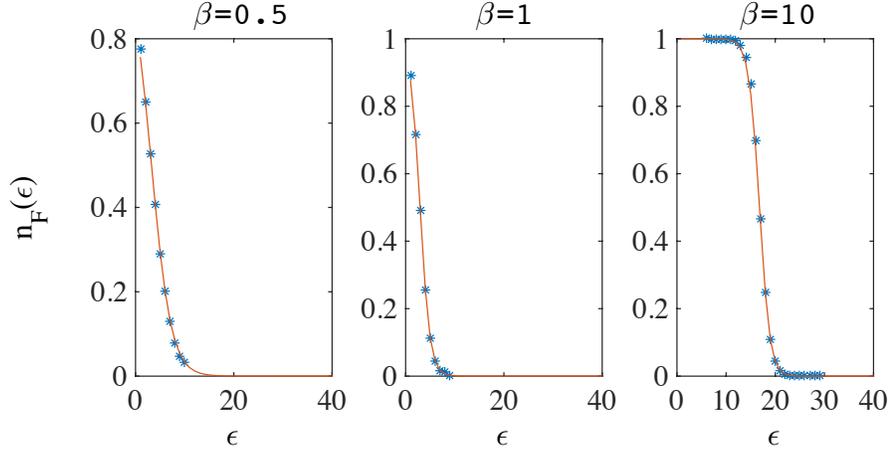}
	\caption{(Color online) The Fermi-Dirac occupation number $n_F(\epsilon)$ is extracted from the simulation results by plotting $\avg{n|\epsilon}/\rho_F(\epsilon)$  (star points) and compared with the theoretical prediction 
	(solid red line). Data are shown for a Fermi-Dirac network with $g(\omega)$ given by a Poisson distribution with  $c=2$. The network size is $N=2000$ and the simulation results are averaged over $30$ runs. }	
	\label{figuredistF}
\end{figure*}

\begin{figure*}
	\includegraphics[width=1.65\columnwidth]{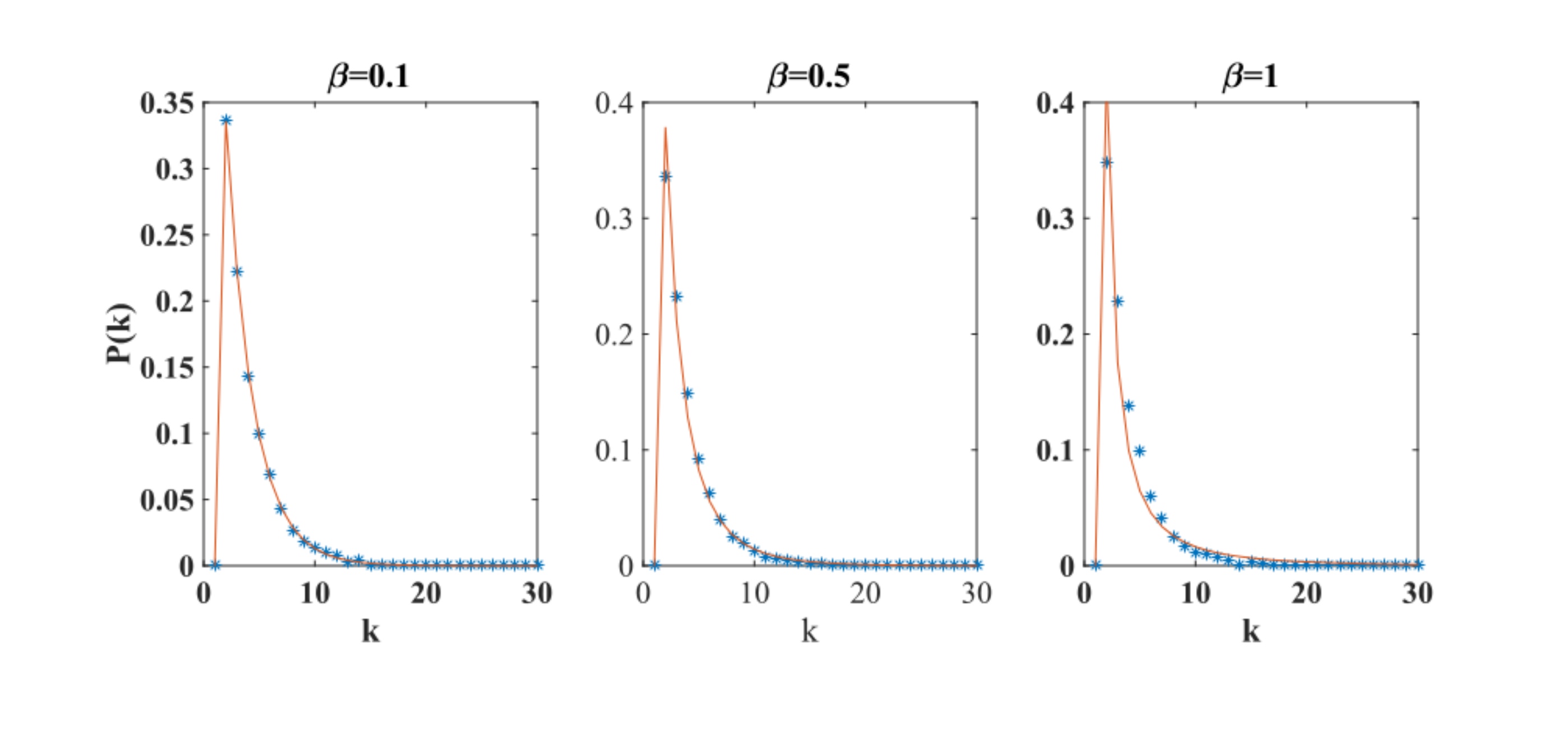}\nonumber \\
	\caption{(Color online) The degree distribution $P(k)$ of the Fermi-Dirac Network is plotted with star points for a Fermi-Dirac network with $g(\omega)$ given by a Poisson distribution with  $c=2$. The network size is $N=5000$ 
	and the simulation results are single realizations of the networks. The simulations are compared with the theoretical predictions of Eq. $(\ref{PkF})$ shown as a solid red line. }
	\label{figuredegreeF}
\end{figure*}

We compare the results of the theory with the outcomes of the simulations as long as the chemical potential $\mu_F$ defined in Eq. $(\ref{selfF})$ is well defined.
In particular we show the results of simulations confirming  that the average of  the  quantum number  $n_{ij}$ over the links of energy $\epsilon$, $\avg{n|\epsilon}$   
is given by  Eq. $(\ref{nFno})$ and follows the Fermi-Dirac statistics.
In Figure $\ref{figuredistF}$ we compare the results of the simulations with the theoretical expectations by plotting
the right and left hand side of equation 
\bea
n_F(\epsilon)=\frac{\avg{n|\epsilon}}{\rho_F(\epsilon)}
\eea
equivalent to Eq. $(\ref{nFno})$
and the  degree distribution $P(k)$ of this network with the theoretical expectation given by Eq. (\ref{PkF}). We find very  good agreement in both cases as displayed in Figure $\ref{figuredegreeF}$.

\subsection{Phase transition in the Fermi-Dirac network}
The self-consistent approach for solving the Fermi-Dirac network is based on the assumption that the chemical potential $\mu_F$ of the network defined in Eq. $(\ref{selfF0})$ exists.
But in the network it is possible to find a phase transition at high enough inverse temperature, i.e. for $\beta>\beta_c$ where this assumption fails (see Sec. \ref{PT}).
In order to determine where this phase transition occurs we have solved the self-consistent equation for the chemical potential $\mu_F$ given by Eq. $(\ref{selfF})$. This equation can
always be solved to find the chemical potential $\mu_F$, but the value of the chemical potential $\mu_F$ as a function of the inverse temperature can have  a maximum for $\beta=\beta_c$.
Here we have identified this maximum with the onset of the phase transition. In fact, from the dynamic rules of the model it is clear that  the network dynamics for increasing value of $\beta$ tends to attach new triangles on unsaturated nodes with lower energy. Therefore if the  probability $P(n=0|\omega,\omega^{\prime})$  is given by Eq. $(\ref{NsF})$   the chemical potential $\mu_F$ that can only increase with increasing inverse temperature  $\beta$.
 
\begin{figure}
\includegraphics[width=0.9\columnwidth]{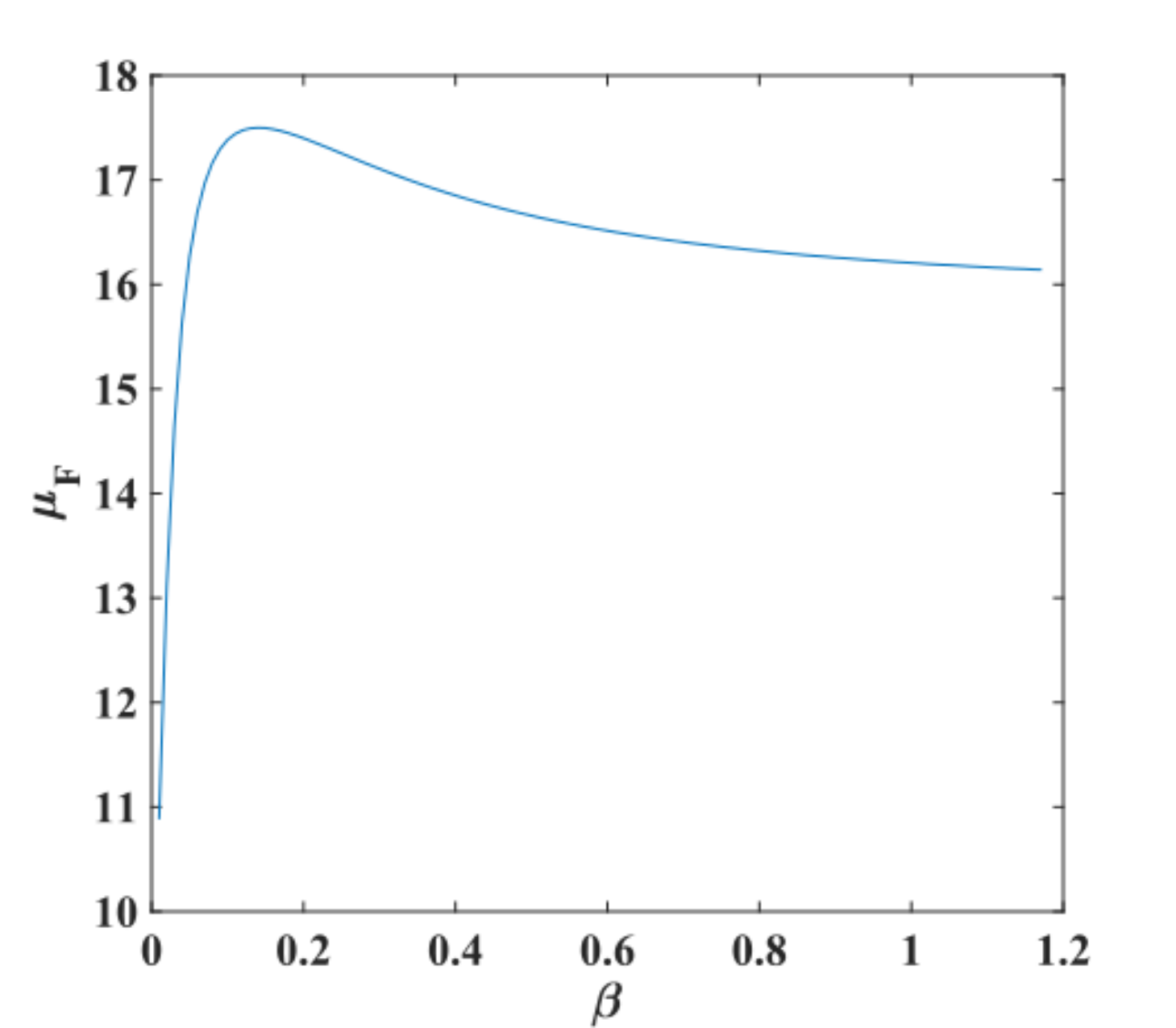}
	\caption{(Color online) The chemical potential $\mu_F$ versus the inverse temperature $\beta$ for a Fermi-Dirac Network with $g(\omega)={\cal N}\frac{1}{\omega!}z^{\omega}$ with $\omega\in [0,100]$ , 
	$c=10$ and  with ${\cal N}$ indicating the normalization sum. The critical value of the inverse temperature is $\beta_c \simeq 0.14$ which   is in good agreement with the simulations 
	(see Figure $(\ref{transition_p0})$).}
	\label{FermiT}
\end{figure}

In  Figure $\ref{FermiT}$ we show the chemical potential $\mu_F$ as a function of the inverse temperature $\beta$ for a Fermi-Dirac network with a Poisson distribution $g(\omega)$ 
(given by $(\ref{Poissondistribution})$) 
with average $c=10$ and linear relation between the energy of the nodes and the energy of the links (Eq. $(\ref{linearenergy})$). In order to perform the numerical calculation of the chemical potential $\mu_F$ 
the distribution $g(\omega)$ is truncated at a cutoff value $\omega_{\Lambda}=100$. The chemical potential $\mu_F$ has a maximum at $\beta_c \simeq 0.14$ which is a good prediction for the 
phase transition as can be seen from the simulation results shown in Figure $\ref{transition_p0}$.

\section{Bose-Einstein Network Evolution}
\subsection{Path integral}
The evolution of the Bose-Einstein quantum state  is described by the  unitary operator $U_t$ defined in the following,
\bea
&&\ket{\psi_N(t)}=U_t\ket{\psi_N(t-1)}\nonumber \\
&&\sqrt{\frac{{\cal Z}(t-1)}{{\cal Z}(t)}}\sum_{\omega_{t+2}}\sum_{i,j|i<j}\sqrt{g(\omega_{t+2})}e^{-\beta \epsilon_{ij}/2}\nonumber \\
&&\times b^{\dag}_{t+2}(\omega_{t+2})c^{\dag}_{(t+2)i}c^{\dag}_{(t+2)j}\tilde{d}^{\dag}_{ij}c^{\dag}_{ij}c_{ij}\ket{\psi_N(t-1)},\eea
where $\epsilon_{ij}=f(\omega_i,\omega_j)$ and  ${\cal Z}={\cal Z}(t)$ is fixed by the normalization condition $\langle{\psi_N(t)}|{\psi_N(t)}\rangle=1$.

In this case the normalization constant ${\cal Z}(t)$ is given by the path integral
\bea\label{BEpartfunc}
{\cal Z}(t)&=&\sum_{\{\omega(t')\}} \sum_{\{\ell(t')\}}  W( \{\omega({t'}),\ell(t')\}_{t' \leq t})
\label{ZBp}
\eea
for $t\geq 2$ and $\epsilon_{ij}=f(\omega_i,\omega_j)$.
Here $\{\omega(t'),\ell_{t'}\}_{t'=1,\ldots, t}$ is a sequence of links $\ell(t')=(i_{t'},j_{t'})$ and a sequence of energies of the new nodes $\omega(t')$ that 
describes a single history over which the path integral is calculated. 
Each path   $\{\omega({t'}),\ell(t')\}_{t' \leq  t}$ in Eq. $(\ref{ZBp})$ is assigned a weight 
 \bea
 W( \{\omega({t'}),\ell(t')\}_{t'\leq t})&=&\prod_{i=1}^{t+2}g(\omega_i)\prod_{t'=2}^t a_{\ell(t')}(t').\nonumber \\
&&\hspace*{-30mm}\times (1+n_{\ell(t')}(t'))e^{-\beta\sum_{i<j}\epsilon_{ij}n_{ij}(t)},
\label{PpB}
 \eea
where
$a_{ij}(t)$ and $n_{ij}(t)$ can be expressed in terms of the history $\{\ell(t')\}_{t'\leq t}$ in the same way as in 
Eq. (\ref{FDaandn}). Note the characteristic sign difference in Eq. (\ref{PpB}) compared to the Fermi-Dirac case.
 Therefore ${\cal Z}(t)$ can be interpreted as a partition function of a statistical mechanics problem in which each path up to time $t$ has probability 
 \bea
 P( \{\omega({t'}),\ell(t')\}_{t'\leq t})=\frac{W( \{\omega({t'}),\ell(t')\}_{t'\leq t})}{{\cal Z}(t)}.
 \label{PpB23}
 \eea

  Each of the  paths $\{\omega({t'}), \ell(t')\}_{t' \leq t}$ can be mapped to a geometrical network evolution with $m=\infty$. 
  In this mapping $\omega(t)$ indicates the energy of the node added to the network at time $t$, $\ell(t)=(i_t,j_t)$ indicates the link to which we attach a new triangle at time $t$, and $a_{ij}(t)$ 
  indicates the adjacency matrix of the network. Moreover $n_{ij}(t)$ indicates the number of triangles that exceed one, attached to the link $(i,j)$.
  The probability of each geometrical network evolution $\{\omega({t'}),\ell(t')\}_{t' \leq t}$ described in Sec. \ref{gmb} is the same as the probability given by  in Eq. $(\ref{PpB23})$.
  
Starting from Eq. $(\ref{PpB})$ we can calculate the conditional probability  that at time $t$ we add a link $\ell(t)=(i,j)$ given the present state of the network evolution,  
$\Pi^{[B]}_{(i,j)}=P(\ell(t)=(i,j)|\{\omega({t'}),\ell(t')\}_{t'<t})$. A straightforward calculation shows that 
\bea
\Pi^{[B]}_{(i,j)}(t)=\frac{e^{-\beta\epsilon_{ij}}a_{ij}(t)(1+n_{ij}(t))}{Z_B},
\label{probB}
\eea
where 
\bea
Z_B=\sum_{i<j}e^{-\beta\epsilon_{ij}}a_{ij}(t)(1+n_{ij}(t)).
\label{ZB}
\eea
The expression in Eq. $(\ref{probF})$ is the same as $\Pi_{(i, j)}^{[1]}$ defined in Eq. $(\ref{prob})$ for $m=\infty$. In this case $\xi_{ij}(t)=1$ for nodes $(i,j)$ for which there is a link, i.e. $a_{ij}(t)=1$.  
It follows that studying the geometrical network evolution for $m=\infty$ determines the properties of the Bose-Einstein quantum network state.
For this reason we call the growing geometrical network with $m=\infty$ the Bose-Einstein Network.

\subsection{ Bose-Einstein statistics}
In the Bose-Einstein Network the average of the quantum number $n_{ij}$ over links with energy $\epsilon$ follows the Bose-Einstein distribution. 
To show this result let us consider the master equation \cite{Doro_book} for the number $N_B^t(n|\omega,\omega')$ of  links $(i,j)$  (with $i>j$ and  $\omega_i=\omega$ and $\omega_j=\omega^{\prime}$) that have  
$n_{ij}=n=0,1,2,\ldots$ at time $t$.
Since at each time we choose a  link $(i,j)$ with probability $\Pi^{[B]}_{(i,j)}$, the master equation reads
\bea
{N_B^{t+1}(n|\omega,\omega')}&=&\frac{e^{-\beta\epsilon }n}{Z_B}N_B^t({n-1}|\omega,\omega')[1-\delta(n,0)]\nonumber \\
&&-\frac{e^{-\beta \epsilon}(n+1)}{Z_B}N_B^t(n|\omega,\omega')\nonumber \\
&&+2\rho_B(\omega,\omega')\delta(n,0)+{N_B^{t}(n|\omega,\omega')},\nonumber\\
\label{NnB}
\eea
where $\epsilon=f(\omega,\omega')$ and 
$\rho_B(\omega,\omega^{\prime})$ is the probability that a new link will connect a new node with energy $\omega$ and an old node with energy $\omega^{\prime}$.
In order to solve this master equation we assume that the normalization constant $Z_B\propto t$ and we put
\bea
e^{-\beta \mu_B}=\lim_{t \to \infty} \frac{Z_B}{t}.
\label{selfB0}
\eea
Moreover we also assume that at large times $N_B^t(n|\omega,\omega') \simeq 2tP_B(n|\omega,\omega')$ as the number of links in the network is $2t+1\simeq 2t $ for $t\gg 1$.
The quantity $P_B(n|\omega,\omega')$ indicates the asymptotic probability that a random link $(i,j)$ with $i>j$ and  $\omega_i=\omega$ and $\omega_j=\omega^{\prime}$,  has $n_{ij}=n$. Making these assumptions, 
it is possible to solve Eq. $(\ref{NnB})$ getting
\bea
P_B(n,|\omega,\omega^{\prime})&=&\rho_B(\omega,\omega^{\prime})\Gamma[1+e^{\beta(\epsilon-{\mu}_B)}]e^{\beta(\epsilon-{\mu}_B)}\nonumber \\ 
&&\times\frac{\Gamma(n+1)}{\Gamma\left[e^{\beta(\epsilon-{\mu}_B)}+n+2\right] }
 \label{NsB}
\eea
for $n\geq 0$.
Eq. $(\ref{NsB})$ is automatically normalized once the distribution $\rho_B(\omega,\omega^{\prime})$ is normalized.
Therefore the average of the quantum numbers $n_{ij}$ over links with energy $\epsilon_{ij}=\epsilon$ is given by 
\bea
\avg{n| \omega,\omega^{\prime}}&=&\rho_B(\omega,\omega^{\prime})\frac{1}{e^{\beta(\epsilon-\mu_B)}-1}\nonumber \\
&=&\rho_B(\omega,\omega^{\prime})n_B(\epsilon),
\label{noop}
\eea
where $n_B(\epsilon)$ indicates the Bose-Einstein occupation number 
\bea
n_B(\epsilon)=\frac{1}{{e^{\beta (\epsilon-{\mu}_B)}-1}}.
\eea
Summing over all the links with energy 
$\epsilon$, we get 
\bea
\avg{n| \epsilon}&=&\rho_B(\epsilon)n_B(\epsilon),
\label{nBno}
\eea
where
\bea
\rho_B(\epsilon)=\sum_{\omega,\omega'}\delta[\epsilon,f(\omega,\omega')]\rho_B(\omega,\omega').
\eea
Therefore in the Bose-Einstein network the average of the quantum number $n_{ij}$ over links of energy $\epsilon$ follows the Bose-Einstein distribution.

To complete the solution  it is necessary to find the correct expression for $\rho_B(\omega,\omega')$.
Since by definition  $\omega$ is the energy of the new node attached to the network at time $t$, and since this energy is drawn randomly from a distribution $g(\omega)$,  the probability 
$\rho_B(\omega,\omega^{\prime})$  can be factorized,
\bea
\rho_B(\omega,\omega')=g(\omega)\tilde{g}_B(\omega^{\prime}),
\label{rggt}\eea
where $\tilde{g}_B(\omega^{\prime})$ is the probability that a new triangle is attached to a link having at its end a node of energy $\omega^{\prime}$ and is therefore normalized.

We can write a recursive equation for $\tilde{g}_B(\omega^{\prime})$. In fact we have 
\bea
\hspace*{-10mm}\tilde{g}_B(\omega)&=&\sum_{\omega^{\prime}}e^{-\beta[f(\omega,\omega')-\mu_B]}\left\{[\rho_B(\omega,\omega^{\prime})+\avg{n|\omega,\omega'}]\right.\nonumber \\
&&\left.+[\rho_B(\omega',\omega)+\avg{n|\omega',\omega}] \right\}\nonumber \\
&=&\sum_{\omega^{\prime}}\left[\rho_B(\omega,\omega^{\prime})+\rho_B(\omega^{\prime},\omega)\right]n_B[f(\omega,\omega')],
\eea
where in the last equation we have substituted Eq. $(\ref{noop})$ into $\avg{n|\omega,\omega^{\prime}}$.
This equation can be formulated as the eigenvalue problem
\bea
\tilde{g}_B(\omega)&=&\sum_{\omega^{\prime}}A_B(\omega,\omega^{\prime})\tilde{g}(\omega^{\prime}),
\eea
where
\bea
A_B(\omega,\omega^{\prime})&=&g(\omega)n_B[f(\omega,\omega^{\prime})]\nonumber \\
&&\hspace*{-5mm}\times \left\{1-\sum_{\omega^{\prime\prime}} g(\omega^{\prime\prime})n_B[f(\omega+\omega^{\prime\prime})]\right\}^{-1}.
\eea
Since we require that $\tilde{g}_B(\omega)$ is a probability, i.e. it is non-negative and  normalized, the solution of the eigenvalue problem is given by the Perron-Frobenious eigenvector $\tilde{g}_B(\omega)$ of the matrix $A_B(\omega,\omega^{\prime})$
satisfying 
\bea
\sum_{\omega}\tilde{g}_B(\omega)=1 .
\eea
By imposing the self-consistent condition in Eq. $(\ref{selfB0})$ we find the equation determining the chemical potential $\mu_B$, 
\bea
\sum_{\epsilon}\rho_B(\epsilon)n_B(\epsilon)=\frac{1}{2},
\label{selfB2}
\eea
which is the same equation as the one fixing the chemical potential in a Bose gas \cite{Kardar} with density of states $\rho_B(\epsilon)$, inverse temperature $\beta$ and specific volume $v=1/2$.

If the   self-consistent Eq. $(\ref{selfB})$ has a solution,  this implies that  the average of the   quantum number $n_{ij}$ over links of energy $\epsilon$, $\avg{n|\epsilon}$, follows the Bose-Einstein  distribution.
\subsection{Structural properties of the Bose-Einstein Network}
In this section we derive the degree distribution and the distribution of the  curvature in the case in which the energies of the links are linearly dependent on the energies of the nodes as in Eq. (\ref{linearenergy}).
In order to derive the degree distribution in this case, let us write the master equation \cite{Doro_book} for the number $N_B^t(k|\omega)$ of nodes that at time $t$ have degree $k$ and energy $\omega$, i.e.
\bea
{ N_B^{t+1}(k|\omega)}&=&\frac{e^{-\beta (\omega-\tilde{\mu}_B)}(k-1)}{t}N_B^t(k-1|\omega)[1-\delta(k,2)]\nonumber \\
&&-\frac{e^{-\beta (\omega -\tilde{\mu}_B)}k}{t}N_B^t(k|\omega)+g(\omega)\delta({k,2})\nonumber \\
&&+{ N_B^{t}(k|\omega)},
\label{NkB}
\eea
where we have assumed self-consistently that asymptotically in time $\tilde{\mu}_B$ is defined as
\bea
e^{\beta \tilde{\mu}_B}=e^{\beta \mu_B}\lim_{t \to \infty} \Avg{\frac{{\sum_{ij}{e^{-\beta \omega_j}a_{ij}(1+n_{ij})}}}{\sum_i k_i \delta(k_i,k)}}.
\label{selfB}
\eea
Assuming that asymptotically in time $N_B^t(k|\omega)\simeq t P(k|\omega)$ we find the expression for $P(k|\omega)$ and substituting this expression in Eq. $(\ref{NkB})$ we have,
\bea
P(k|\omega)= \frac{g(\omega)\Gamma[2+e^{\beta(\omega-\tilde{\mu}_B)}]e^{\beta(\omega-\tilde{\mu}_B)}\Gamma(k)}{\Gamma\left[e^{\beta(\omega-\tilde{\mu}_B)}+k+1\right] }
\eea
for $k\geq 2$.
Therefore the degree distribution of the entire network is scale-free and  given by 
\bea
P(k)&=&\sum_{\omega}P(k|\omega)\nonumber \\
\label{PkB}
\eea
for $k\geq 2$.
The fraction of the total number  of links attached to nodes  with energy  $\omega$ is given, asymptotically in time, by 
\bea
\avg{k|\omega}=\frac{1}{2}\sum_{k}{k}P(k|\omega)= g(\omega)  [1+\tilde{n}_B(\omega)],
\label{k0}
\eea
where $\tilde{n}_B(\omega)$ is defined as 
\bea
\tilde{n}_B(\omega)=\frac{1}{e^{\beta (\omega-\tilde{\mu}_B)}-1}.
\eea
In Eq. $(\ref{k0})$ the first term indicates the fraction of  links initially attached to the new nodes of energy $\omega$, and is therefore  given by $g(\omega)$ because every new link has one end attached to a new node and the new node has energy $\omega$ with probability $g(\omega)$.
The second term represents the fraction of links attached to the nodes of energy $\omega$, after the time of their arrival into the network. This term is proportional to the Bose-Einstein occupation number.

The curvature $R_i$ of a node $i$ is given by 
\bea
R_i=\frac{4-k_i}{6}.
\eea
So the distribution of the curvature $P(R)$ is given by 
\bea
P(R)=\sum_{\omega} \frac{g(\omega)\Gamma[2+e^{\beta(\omega-\tilde{\mu}_B)}]e^{\beta(\omega-\tilde{\mu}_B)}\Gamma(4-6R)}{\Gamma\left[e^{\beta(\omega-\tilde{\mu}_B)}+5-6R\right] }
\eea
where $R\leq \frac{1}{3}$. Therefore the distribution of the curvature is scale-free and decaying as a power-law for  negative values of the  curvature.Moreover the average curvature is $\Avg{R}=1/N$ and the fluctuations around this average  $N\to \infty $ are diverging, i.e. $\Avg{R^2}\to \infty$ as $N\to \infty$.

 \subsection{Comparison with numerical simulations}
\begin{figure*}
	\includegraphics[width=1.6\columnwidth]{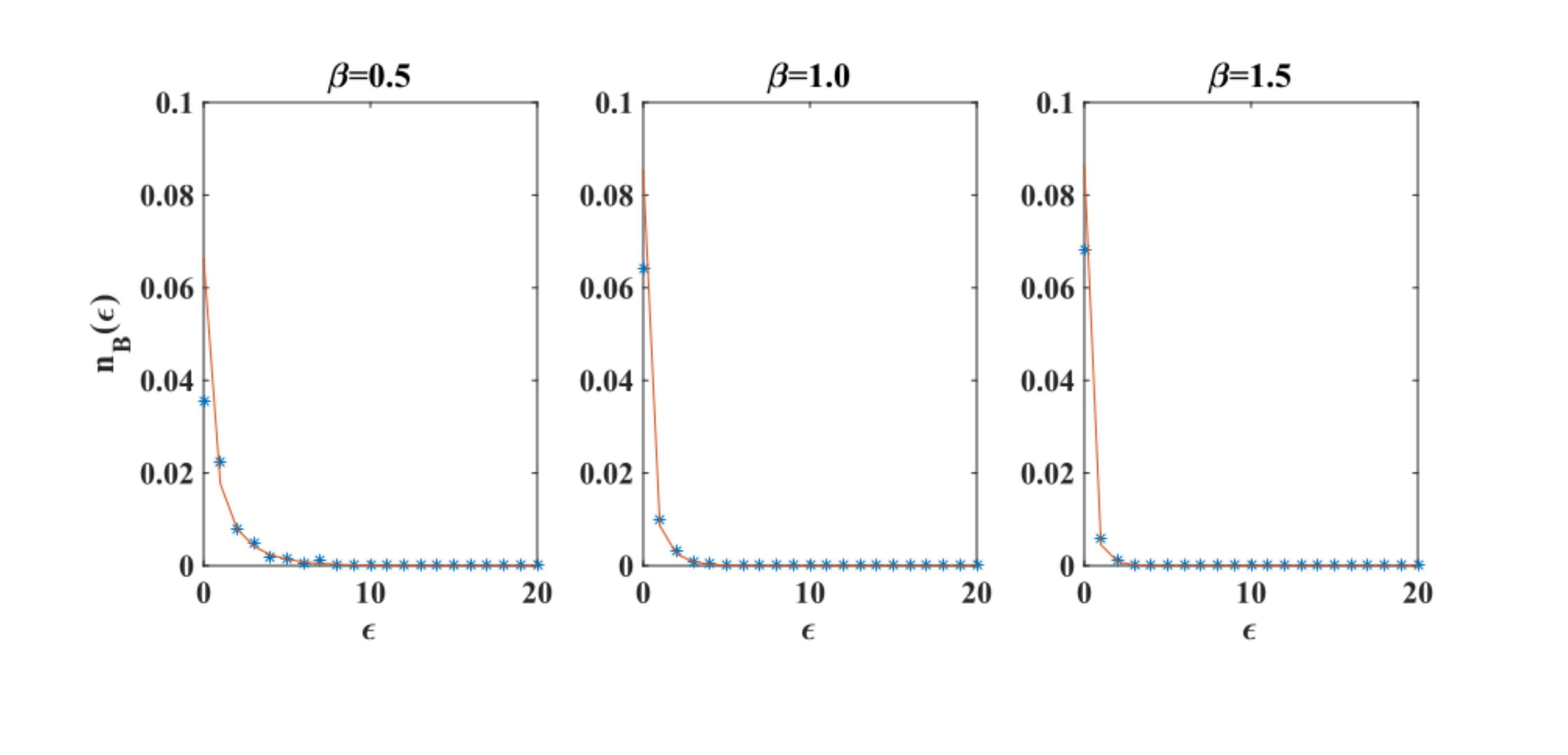}
	\caption{(Color online) The Bose-Einstein occupation number $n_B(\epsilon)$ is extracted from the simulation results by plotting $\avg{n|\epsilon}/\rho_B(\epsilon)$  (star points) and compared with the theoretical prediction 
	(solid red line). Data are shown for a Fermi-Dirac network with $g(\omega)$ given by a Poisson distribution with  $c=2$. The network size is $N=1000$ and the simulation results are averaged over $50$ runs. }	
	\label{figuredistB}
\end{figure*}

\begin{figure*}
	\includegraphics[width=01.6\columnwidth]{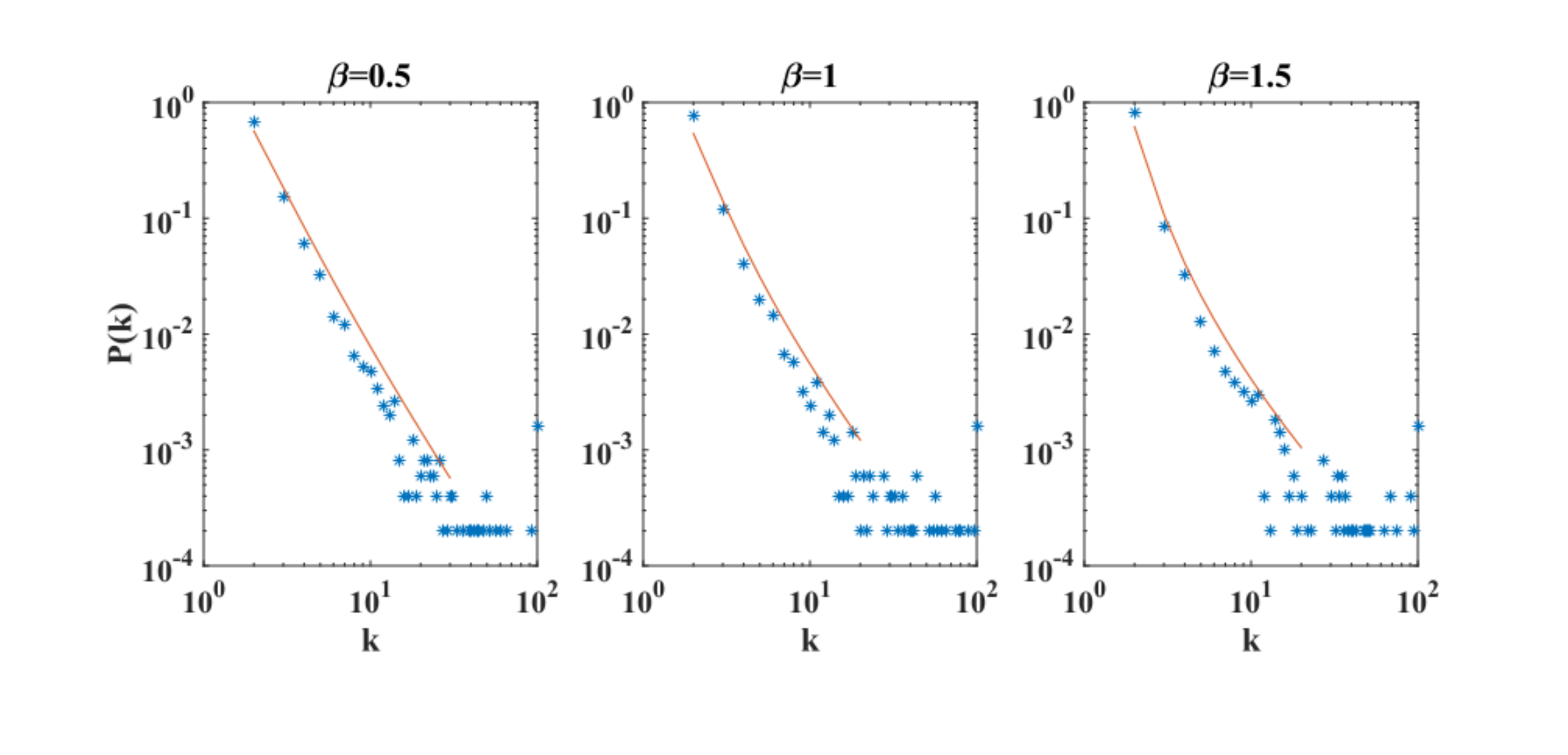}
		\caption{(Color online) The degree distribution $P(k)$ of the Bose-Einstein Network is plotted with star points for a Bose-Einstein  Network with $g(\omega)$ given by a Poisson distribution with  $c=2$. 
		The network size is $N=5000$ and the simulation results correspond to a single network realization. The simulations are compared with the theoretical predictions of Eq. $(\ref{PkB})$ shown as a solid red line. }	
	\label{figuredegreeB}
\end{figure*}

We numerically simulate a Bose-Einstein Network Evolution in which the energies of the nodes are non-negative integers with $g(\omega)$ given by a Poisson distribution with average $c$ 
(Eq. (\ref{Poissondistribution})) and the link and node energies are related by Eq. (\ref{linearenergy})
and compare with the theoretical results.
We  verify that Eq. $(\ref{nBno})$
is satisfied by the results of the numerical simulations.
In Figure \ref{figuredistB} we plot both sides of the equation 
\bea
n_B(\omega)=\frac{\avg{n|\epsilon}}{\rho_B(\epsilon)},
\eea
equivalent to Eq. ($\ref{nBno}$)
and the degree distribution $P(k)$ of this network with the theoretical expectation given by Eq. $(\ref{PkB})$,  finding very good agreement between theoretical and numerical
results in both cases (see Figure $\ref{figuredegreeB}$).

\subsection{Transition: Bose-Einstein condensation}
\begin{figure}
\includegraphics[width=0.9\columnwidth]{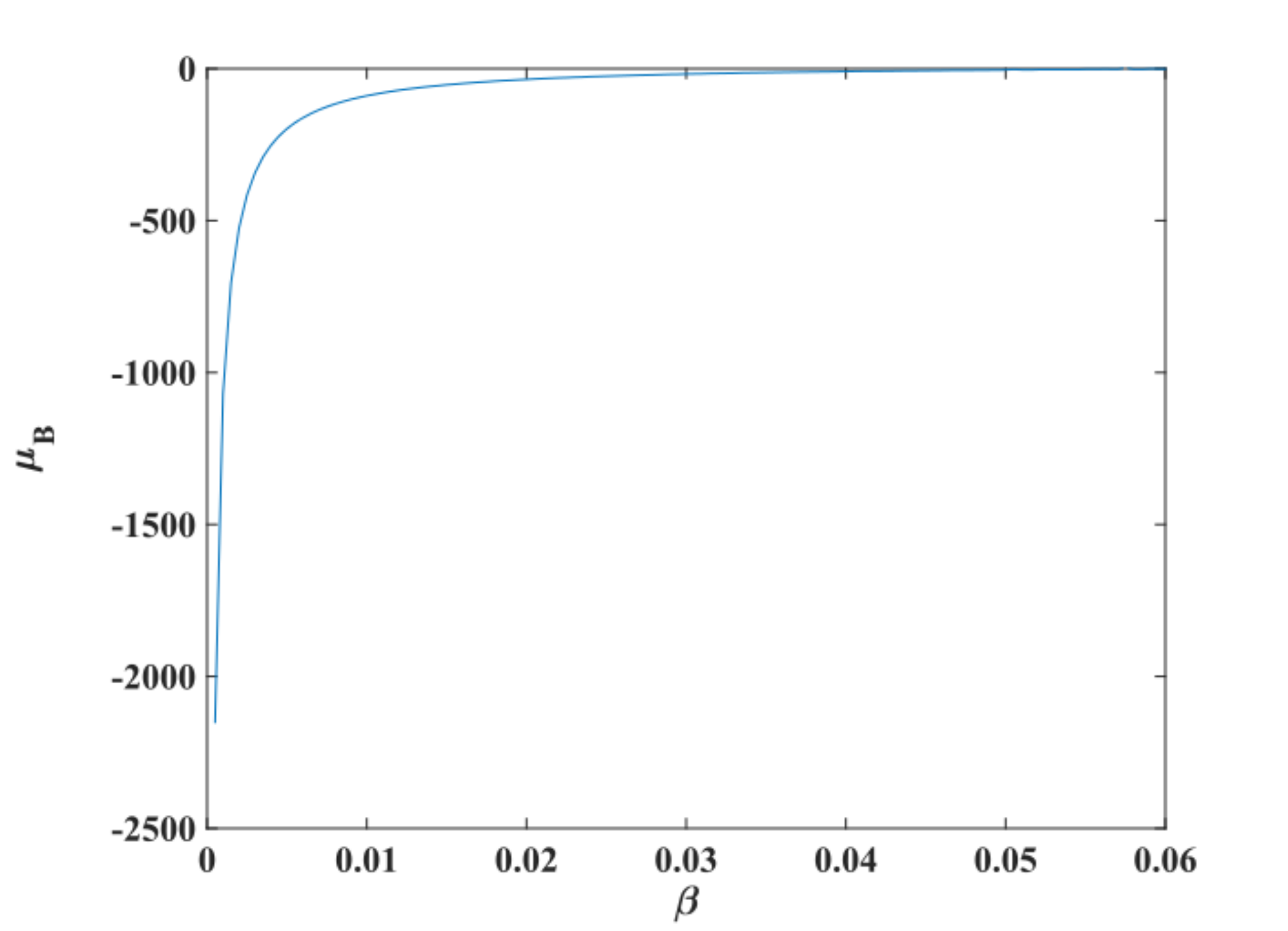}
	\caption{(Color online) The chemical potential $\mu_B$ versus the inverse temperature $\beta$ for a Bose-Einstein Network with $g(\omega)={\cal N}\frac{1}{\omega!}z^{\omega}$ with $\omega\in [0,100]$, $c=10$ and ${\cal N}$ indicating a normalization sum. The critical value of the inverse temperature is $\beta_c \simeq 0.06$  which is in good agreement with the simulations (see Figure $\ref{transition_p0}$).}
	\label{BoseT}
\end{figure}

The self-consistent approach for solving the Bose-Einstein network is based on the assumption that the chemical potential $\mu_B$ of the network defined in Eq. $(\ref{selfB0})$ exists.
But in the network it is possible to find a phase transition at high enough inverse temperature, i.e. for $\beta>\beta_c$, where this assumption fails.
Since the negative chemical potential, $\mu_B<0$, can only increase with the temperature, the critical value of the inverse temperature $\beta_c$ is determined by the self-consistent equation for the 
chemical potential Eq. $(\ref{selfB2})$ where we impose $\mu_B=0$ and $\beta=\beta_c$, i.e.
\bea
\sum_{\epsilon}\rho_B^c(\epsilon)\frac{1}{e^{\beta_c\epsilon}-1}=\frac{1}{2}\ .
\eea 
Here $\rho_B^c(\epsilon)$ is given by Eq.$(\ref{rggt})$ and is calculated for $\mu_B=0$.
In a Bose Gas of density of states $\rho_B(\epsilon)$ the existence of a finite critical temperature $\beta_c$ indicates  the onset of the Bose-Einstein condensation. 
For the network this indicates that there is a single link incident to a finite fraction of all the triangles, and therefore also the degree of the incident nodes is a finite fraction of the total degree of the network 
(see Figures (\ref{visualization_m1}), (\ref{transition_p0}), (\ref{transition_time_p0}) and the discussion of the transition in Sec. \ref{PT}).
This phenomenon is similar to the one  occurring in other models displaying the so-called Bose-Einstein condensation in complex networks \cite{Bose,Weighted}.
In Figure $\ref{BoseT}$ we show the chemical potential $\mu_B$ as a function of the inverse temperature $\beta$ for a Bose-Einstein network with a Poisson distribution $g(\omega)$ with average $c=10$ 
(given by Eq. $(\ref{Poissondistribution})$) and a linear relation between the energy of the links and the nodes (given by Eq. $(\ref{linearenergy})$), where in order to perform the numerical calculation of 
$\mu_B$ the distribution 
$g(\omega)$ is truncated at a cutoff value $\omega_{\Lambda}=100$. In this network a Bose-Einstein phase transition occurs at $\beta_c\simeq 0.06$ which is in very good agreement with the simulation results shown in Figure 
$\ref{transition_p0}$.
\section{Thermodynamics of  quantum geometric networks}
\subsection{Relation between the total energy $E$ and the entropy $S$}
Given the quantum geometric network evolution it is natural to characterize its thermodynamic properties above and below the phase transition.
Let us define the total energy $E$ of a  quantum geometric network as 
\bea
E(t)=\sum_{i<j}\epsilon_{ij} n_{ij}(t)
\label{energy}
\eea
and the entropy $S(t)$ of the quantum geometric  network evolution as 
\bea
S(t)&=&-\sum_{\{\ell(t)\}_{t'\leq t}}P(\{\ell(t')\}_{t' \leq t}|\{\omega(t')\}_{t'<t})\nonumber\\
&&\ln P(\{\ell(t')\}_{t'\leq t}|\{\omega(t')\}_{t'<t}).
\eea
In this expression, $P(\{\ell(t')\}_{t' \leq t}|\{\omega(t')\}_{t'< t})$ is the probability that  the temporal evolution  of the network until time $t$ is described by the subsequent addition of triangles to the links  $\{\ell(t')\}_{t' \leq t}$,  given that the energies 
of the nodes until time $t-1$ are  $\{\omega(t')\}_{t'<t}$.
Together with the definition of  $H^{[1]}$ given by Eq. ($\ref{H1}$) it can be easily derived that 
\bea
H^{[1]}(t)=\Delta S(t)=S(t)-S(t-1).
\label{HS}
\eea
Moreover we have already found (Eq. $(\ref{H1eZ})$) that 
\bea
H^{[1]}=\beta \Avg{\epsilon_{ij}}+\ln Z,
\label{HZ}
\eea
where  the average $\Avg{\epsilon_{ij}}$ is given by Eq. $(\ref{ep})$
and can be related to  the expected increment in time of the  energy $E(t)$ given by Eq. $(\ref{energy})$,  
\bea
\Avg{\epsilon_{ij}}=&\Avg{\Delta E}=\Avg{E(t)-E(t-1)}.
\label{DE}
\eea
The relation between $\Delta S$ and $\Avg{\Delta E}$ can be found using Eqs. $(\ref{HS})-(\ref{HZ})$ and $(\ref{DE})$.  This relation depends on the inverse temperature $\beta$.
For $\beta<\beta_c$ we have that for large times, $t\gg 1$, $Z\simeq e^{-\beta \mu}t$, and therefore 
\bea
\Delta S\simeq \beta \left[\Avg{\Delta E}-\mu\right]+\ln t.
\eea
In the limit $t\to \infty$, the chemical potential  $\mu$ converges to $\mu_F$ for the Fermi-Dirac network and to $\mu_B$ for the Bose-Einstein network.
Instead, for $\beta>\beta_c$ we have that $Z={\cal O}(1)$.  By putting $Z=e^{-\beta \nu}={\cal O}(1)$, we have 
\bea
\Delta S=\beta \left[\Avg{\Delta E}-\nu \right]
\eea
where $\nu$ is a stochastic variable depending on the history of the network.

\subsection{Probability of an unlabeled quantum  network final state}
Given a  network quantum state mapped to a geometric network with $N=t+2$ nodes $i=1,2,\ldots, N$,
\bea
\ket{\phi_L(t)}=\prod_i \ket{\omega_{i}}\prod_{i<j} \ket{a_{ij}}\prod_{i<j|a_{ij}=1}\ket{n_{ij}},
\eea
consider the unlabelled  quantum network state constructed from it by considering all the permutations of the node labels
\bea
\ket{\phi_S(t)}&=&\sum_{\{\pi(i)\}}\prod_i \ket{\omega_{\pi(i)}}\prod_{i<j} \ket{a_{\pi(i)\pi(j)}}\nonumber \\
&&\times \prod_{i<j|a_{\pi(i)\pi(j)}=1}\ket{n_{\pi(i)\pi(j)}},
\eea
where $\pi(i)$ indicates a permutation of all the indices $\{i=1,2\ldots, N\}$ of the nodes.
We are interested in evaluating the probability $P_f$ that the quantum network state $\ket{\psi_N(t)}$ given Eq. $(\ref{psiNt})$
is found in this unlabelled final network state, i.e. we want to calculate 
\bea
P_f=|\bra{\phi_S(t)}\psi_N(t)\rangle|^2.
\eea 
This is equivalent to calculate the probability that the geometrical network evolution generate a time $t$  networks that are equivalent under relabeling to the nodes. Any given history from time $t=1$ to time $t$, corresponds to a series of processes consisting in gluing new triangles to unsaturated links. Therefore, given a final network state one can prune the network in the same order at which the triangles have been added. In this pruning process, we start by removing the last triangle it has been glued to the network, then removing the second last triangle until we reach the initial triangle of the network evolution.
Nevertheless, different network evolutions can generate networks that only differ by a relabeling of the nodes corresponding to the same set of triangles added in a different order starting from the initial triangle.
Finding how many such histories exist is a  problem can be cast into the problem of finding the number {\cal N} of  ways we can prune the network corresponding to the final unlabelled network state.
The pruning of a given geometric  network  consists in an iterative process in which one takes randomly any triangle incident only to a single other triangle and 
different from the initial triangle, and removes it, until the full network is reduced to the single initial triangle.
If we call ${\cal N}$ the number of ways this pruning can be done, we find that 
\bea
P_f=\frac{{\cal N}^2\left(\prod_{i<j|a_{ij}=1}n_{ij}!\right)}{\cal Z}e^{-\beta E}
\eea
where  $E$ is given by Eq. $(\ref{energy})$.
Finding ${\cal N}$ is a combinatorial problem to be solved for any given final network realization. The formula for ${\cal N}$ can be found iteratively using similar techniques as in \cite{Burda1,Burda2}.
We can interpret  
\bea
{\cal S}=2\log({\cal N})
\eea
 as an entropy associated to the unlabelled network state.
With this notation we have for the Fermi-Dirac Network,
\bea
P_f=\frac{e^{-\beta E+{\cal S}}}{\cal Z},
\eea
and for the Bose-Einstein Network
\bea
P_f=\frac{e^{-\beta E+{\cal S}}\left(\prod_{i<j|a_{ij}=1}n_{ij}!\right)}{\cal Z},
\eea

\section{Generalized geometric network model with energy of the links}
\subsection{The network evolution}
Here we consider an extension of the geometric network  model that might allow the generation of networks in which not all the nodes are at the boundary. In particular our goal is to describe geometric network models in which the resulting networks might contain saturated nodes, i.e. nodes incident only to saturated links. 

Therefore here we define a  generalized geometric network model where the  non equilibrium  network evolution  includes and additional  processes with respect to the one introduced for the geometric networks studied in the previous sections.   
We start from a network formed by a single triangle,  a simplex of dimension $d_n=2$. 
Each link can belong at most to $m$ triangles. If a link $(i,j)$ belongs to $m$ triangles it is saturated and $\xi_{ij}=0$. If it belongs to less than $m$ triangles it is unsaturated and $\xi_{ij}=1$.
To each node $i$ an energy of the node $\omega_i$ is assigned from a distribution $g(\omega)$. The energy of the node is quenched and does not change during the evolution of the network.
Moreover to each link $(i,j)$ we associate an energy of the link $\epsilon_{ij}$ given by a symmetric function of the energy of the nodes $i$ and $j$ as in Eq. (\ref{unob}).

At each time we perform two processes: process (a) and process (b).
The process (a) is the same process considered in the original model described in Sec. \ref{gmb}. Here we consider also an additional process (process (b)) occurring at each time with probability $p$.
Process (a) and process (b) are described in the following.
\begin{itemize}
\item{\em Process (a)-} We add a triangle to an unsaturated link $(i, j)$ of the network linking node $i$ to node $j$. We choose this link with   probability  $\Pi_{(i, j)}^{[1]}$ given by  Eq. $(\ref{prob})$.
Having chosen the link $(i, j)$ we add a  node $r$,  two links $(i, r)$ and $(j, r)$ and the new triangle linking node $i$,  node $j$ and node $r$.

\item{\em Process (b)-}	
With probability $p$ we add a single link between two nodes not already linked and at hopping distance $2$,  and we add all the triangles that this link closes,  without adding more than $m$ triangles to each link. 
In order to do this,  we define a variable $\sigma(t)=1$ if process (b) takes place at time $t$ (event which occurs with probability $p$) and $\sigma(t)=0$ if process (b) does not take place at time $t$.
If $\sigma(t)=1$ we choose two unsaturated  links $(i', j')$ and $(j',r')$  specified by ${\bf q}=(i',j',r')$ with probability $\Pi_{{\bf q}=(i',j',r')}^{[2]}$ given by 
\bea
\Pi^{[2]}_{{\bf q}=(i',j',r')}&=&\frac{1}{\cal C}e^{-\beta (\epsilon_{i'j'}+\epsilon_{j'r'})}(1-a_{i'r'})\nonumber \\
&&\hspace{-15mm}\times a_{i'j'}\xi_{i'j'}a_{j'r'}\xi_{j'r'}(1+n_{i'j'})(1+n_{j'r'})\nonumber \\
&&\hspace{-15mm}\times\prod_{s\neq r'}\left(a_{i's}\xi_{i's}a_{sr'}\xi_{sr'}(1+n_{i's})(1+n_{sr'})\right)
\label{Pq}
\eea
where  ${\cal C}$ is the normalization constant and where $\xi_{ij}=1$ indicates an unsaturated link, while $\xi_{ij}=0$ indicates a saturated link. Moreover in Eq. $(\ref{Pq})$ the quantity $1+n_{ij}$ indicates the total number of triangles incident to the link $(i,j)$.
Then we add a link $(i', r')$ and all the triangles  that the link $(i',r')$ closes.
\end{itemize}
\begin{figure}
	\includegraphics[width=0.95\columnwidth]{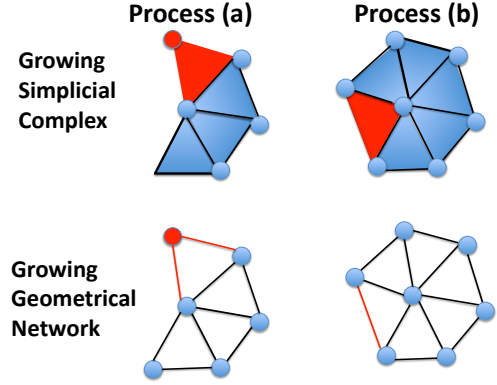}\nonumber \\
	\caption{(Color online) The generalized growing geometric network model is the underlying network of a growing simplicial complex evolving by process (a) and process (b). In process (a) a new triangle is connected to the network and 
	glued to an existing unsaturated link of the network. In process (b) two nodes at distance $2$ connected by unsaturated links are  connected by a new link, and all the triangles that this link closes are added, provided that no more than $m$ triangles are incident to each link. In the figure the case with $m=2$ is plotted.}	
	\label{figuresimg}
\end{figure}
In Figure $\ref{figuresimg}$ we show how the generalized growing network model can be extracted by the generalized growing simplicial complex evolving through process (a) and process (b).

\begin{figure}
$\begin{array}{c}
	\includegraphics[width=0.95\columnwidth]{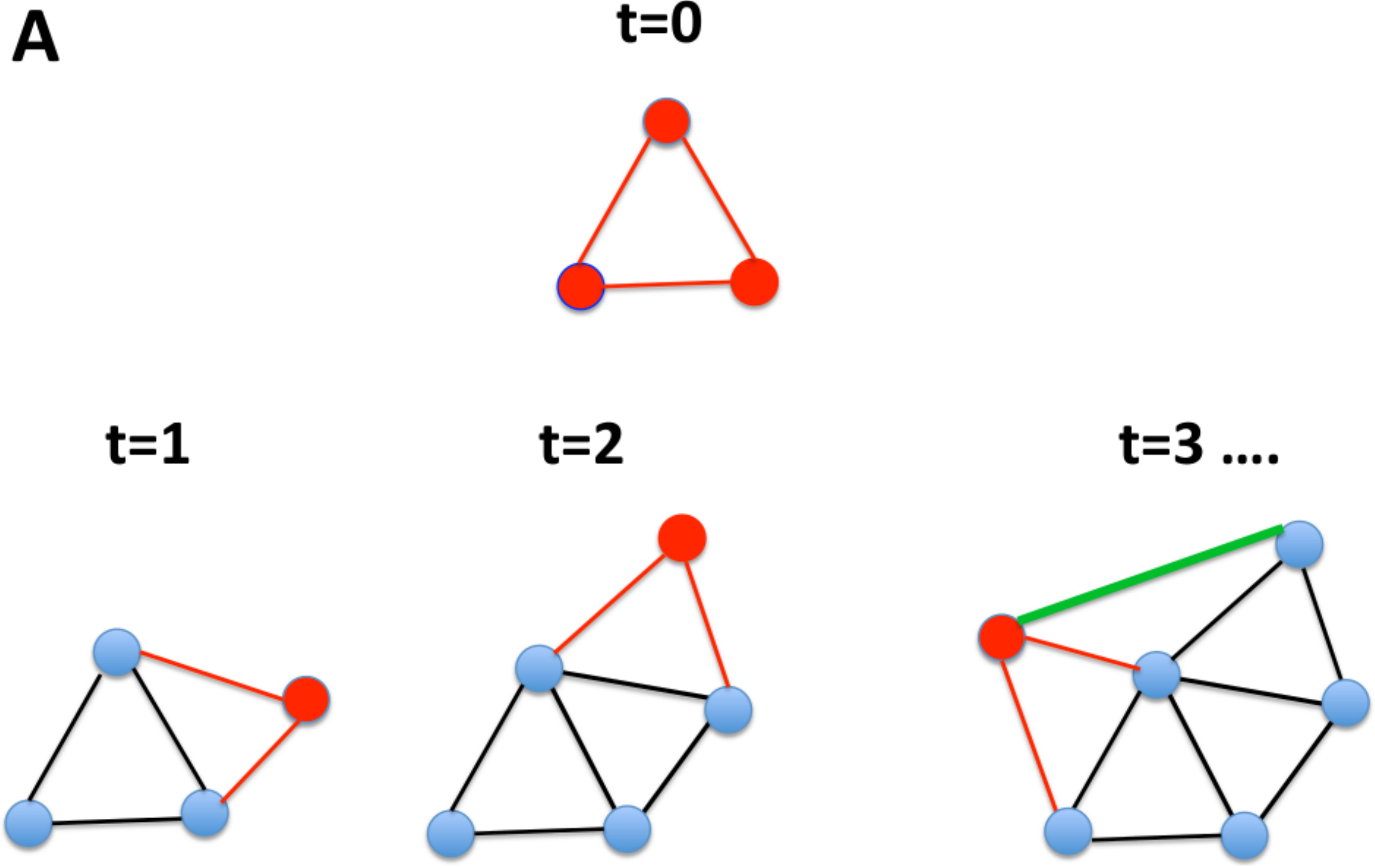}\nonumber \\
	\includegraphics[width=0.95\columnwidth]{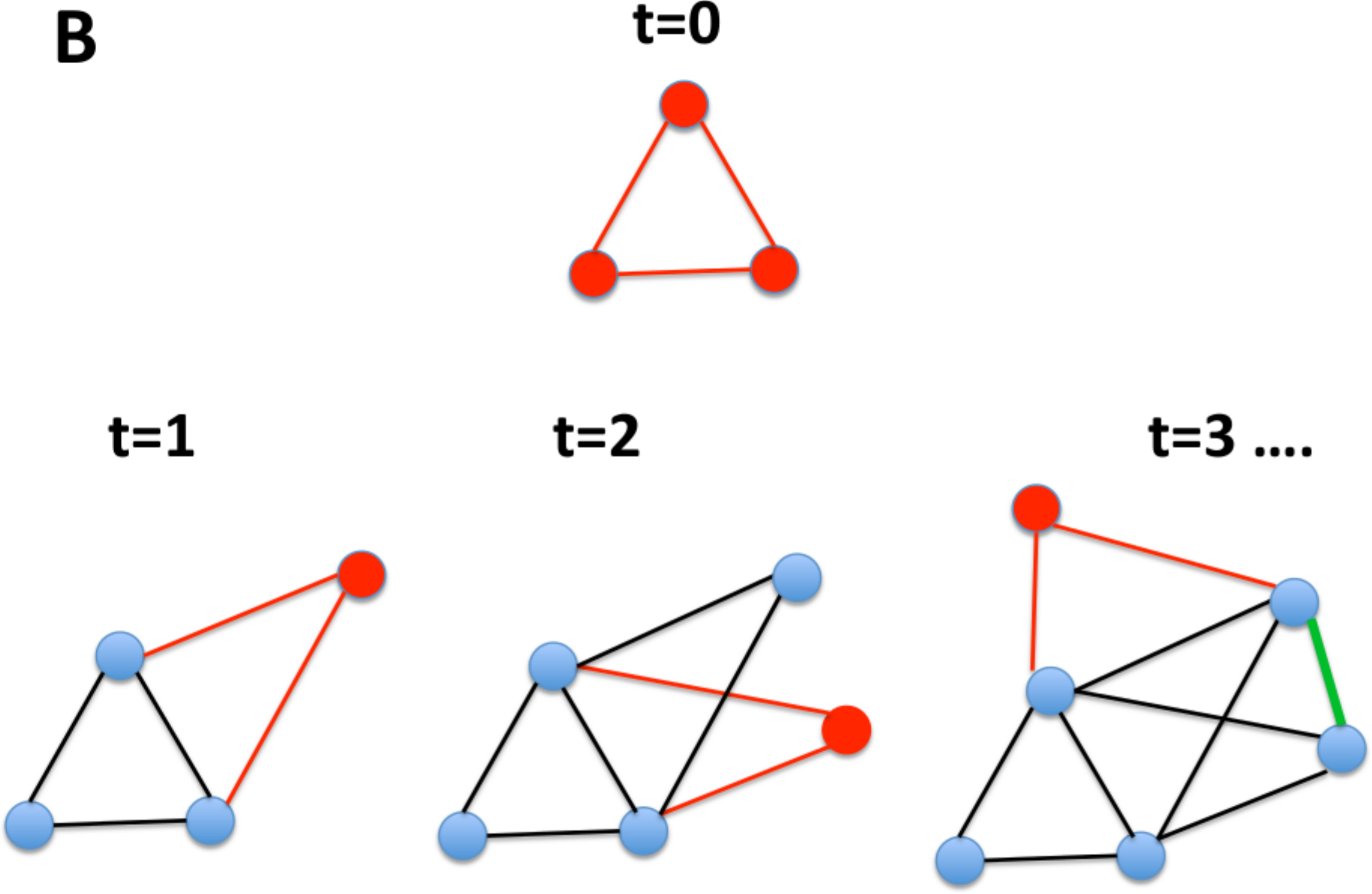}\nonumber \\
	\end{array}$
	\caption{The Generalized Fermi-Dirac Network evolution (panel A) and the Generalized Bose-Einstein Network Evolution (panel B). At each time a new triangle is added to a link $(i,j)$ chosen according to 
	the probability $\Pi_{(i, j)}^{[1]}$ 
	given by Eq. $(\ref{prob})$ (process (a)), and with probability $p$ also process (b) takes place and a new link is added between two nodes at distance 2 with probability chosen with 
	$\Pi^{[2]}_{{\bf q}(t)}$ given by 
	Eq. (\ref{Pq}). The maximal number of triangles incident to a link are $m=2$ for the Generalized Fermi-Dirac Network Evolution and $m=\infty$ for the Generalized Bose-Einstein Network evolution.  In the figure, at time $t=3$ the  
	process $(b)$ takes place and the new link added according to this process is plotted with a thick green line. }	
	\label{figure34}
\end{figure}

In Figure \ref{figure34} we show schematically the dynamical rules for building  generalized geometrical growing networks with $m=2$ and $m=\infty$ that we will call respectively 
Generalized Fermi-Dirac Network  and Generalized Bose-Einstein Network.

\begin{figure*}
	\includegraphics[width=1.9\columnwidth]{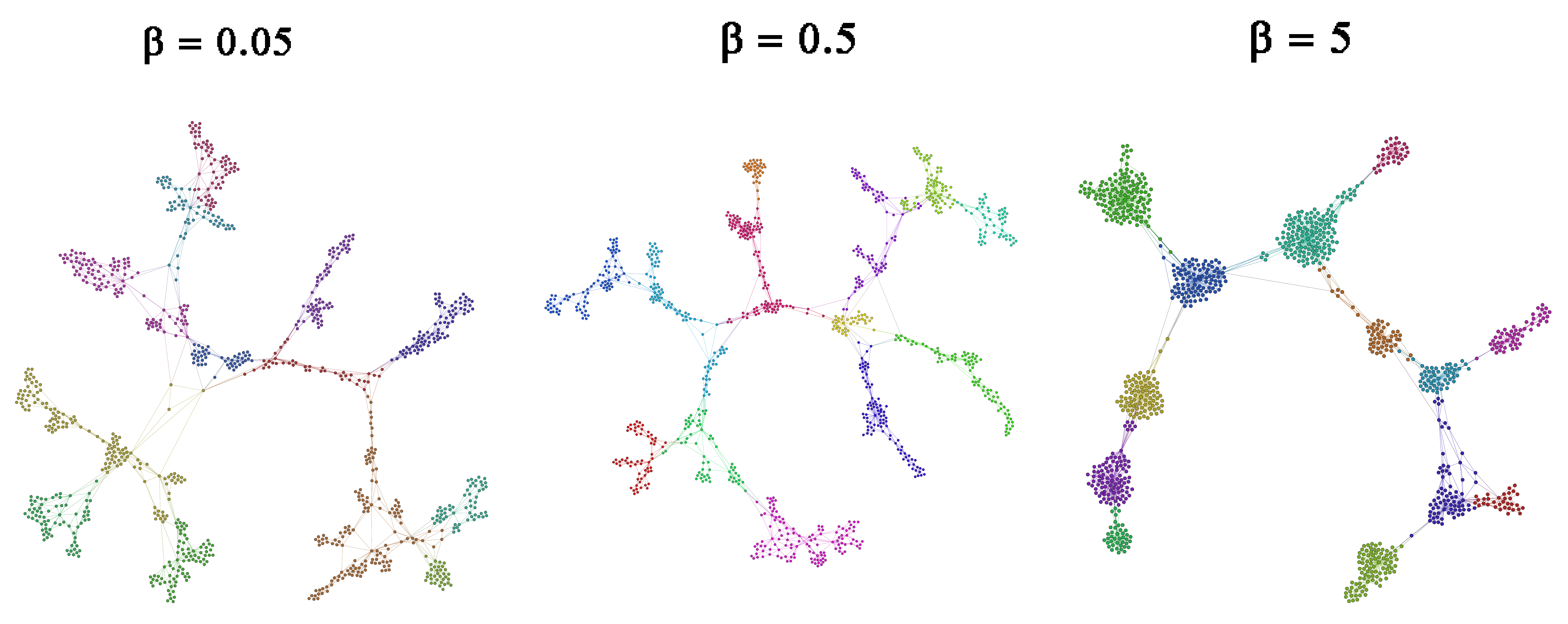}
	\caption{(Color online) Visualization of the  Generalized Fermi-Dirac Network  with a Poisson energy distribution of the nodes $g(\omega)$ given by Eq. $(\ref{Poissondistribution})$,   $c=10$, 
	$\beta=0.05,0.5,5$, $N=1000$ and $p=0.9$. For low value of $\beta$, i.e. $\beta<\beta_c$, the network is small-world, for large values of $\beta$, i.e. $\beta>\beta_c$, 
	the network develops a large diameter. The color indicates the partition into communities found by running the Louvain algorithm \cite{Louvain}.}	
	\label{visualization_m2b}
\end{figure*}
\begin{figure*}
\includegraphics[width=1.9\columnwidth]{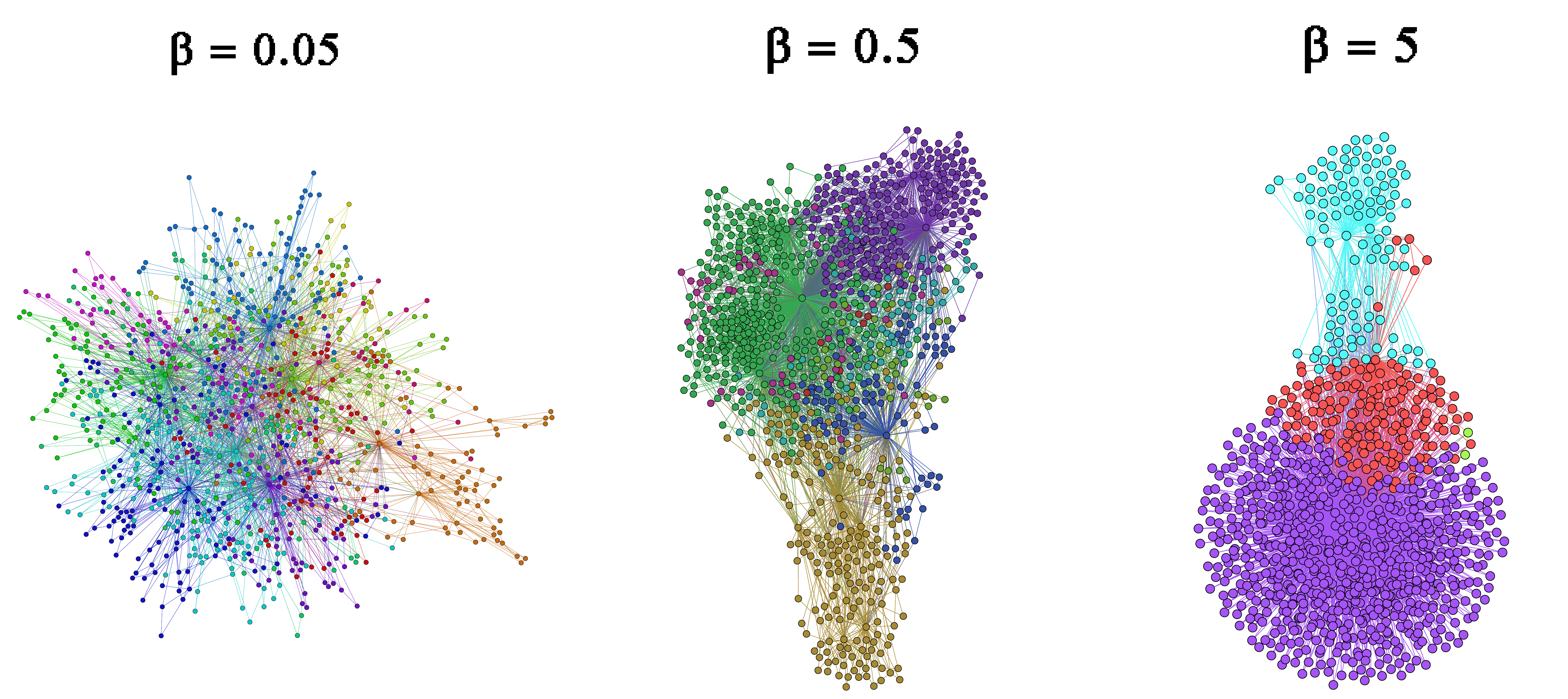}
	\caption{(Color online) Visualization of the  Generalized Bose-Einstein Network  with a Poisson energy distribution of the nodes $g(\omega)$ given by Eq. $(\ref{Poissondistribution})$,   
	$c=10$, $\beta=0.05,0.5,5$,$N=1000$ and $p=0.9$. For low value of $\beta$, i.e. $\beta<\beta_c$, the network is small-world, for large values of $\beta$, i.e. $\beta>\beta_c$, 
	the network is condensed, and develops a small diameter. The colour indicates the partition into communities found by running the Louvain algorithm \cite{Louvain}.}
	\label{visualization_m1b}
\end{figure*}
\subsection{Entropy rate of the Generalized Geometric Network}
Calling $\Omega(t)=\{\omega(t),\ell(t),\sigma(t),{\bf q}(t)\}$
the generalized geometric network evolution is described by the sequence $\{\Omega(t')\}_{t'\leq t}$.
At time $t$ the sequence includes the new terms $\Omega(t)$.
The entropy rate of the generalized geometric growing network is 
\bea
H_G(t)&=&-\sum_{\Omega(t)}\sum_{{\bf q}(t)}{\cal P}(\Omega(t)|\{\Omega(t')\}_{t'<t})\nonumber \\
&&\times\ln {\cal P}(\Omega(t)|\{\Omega(t')\}_{t'<t}))\ ,
\eea
where ${\cal P}(\Omega(t)|\{\Omega(t')\}_{t'<t}))$ indicates the probability of having $\Omega(t)$ given $\{\Omega(t')\}_{t'<t}$.
The probability of $\omega(t)=\omega$ is $g(\omega)$, and it is independent of all the other precedent events.
The probability of $\ell(t)$ is given by $\Pi^{[1]}_{\ell(t)}$ as in Eq. $(\ref{prob})$.
The probability of having process (b), i.e. $\sigma(t)=1$, is $p$, and the probability of not having process (b), i.e. $\sigma(t)=0$, is $1-p$.
Finally if the process (b) takes place, the probability of ${\bf q}(t)$ is $\Pi^{[2]}_{{\bf q}(t)}$ as in Eq. ($\ref{Pq}$).
The entropy rate of the network evolution can therefore be written as the sum of three different entropy rates 
\bea
H_{G}(t)=H_{\omega}+H^{[1]}(t)+H^{[2]}(t).
\eea
Here, $H_{\omega}$ is the contribution to the  entropy rate due to the time-independent random distribution of the energy of the nodes, 
\bea
H_{\omega}=-\sum_{\omega} g(\omega)\ln g(\omega).
\eea
The quantity $H^{[1]}(t)$ indicates the contribution due to process (a),
\bea
H^{[1]}(t)=-\sum_{\ell(t)}\Pi^{[1]}_{\ell(t)}\ln \Pi^{[1]}_{\ell(t)},
\eea
while $H^{[2]}(t)$ indicates the contribution due to process $(b)$, 
\bea
\hspace*{-5mm}H^{[2]}(t)=-(1-p)\ln(1-p)-\sum_{{\bf q}(t)}p\Pi^{[2]}_{{\bf q}(t)}\ln \left[p\Pi^{[2]}_{{\bf q}(t)}\right].
\eea
 A change in the scaling of $H^{[1]}(t)+H^{[2]}(t)$ with time indicates a phase transition in the network. Such a phase transition can occur at high values of the inverse temperature $\beta$, where the network dynamics can become extremal, 
 similar to what we have seen occurring in the precedent sections in the case $p=0$.
\begin{figure*}	
\includegraphics[width=1.5\columnwidth]{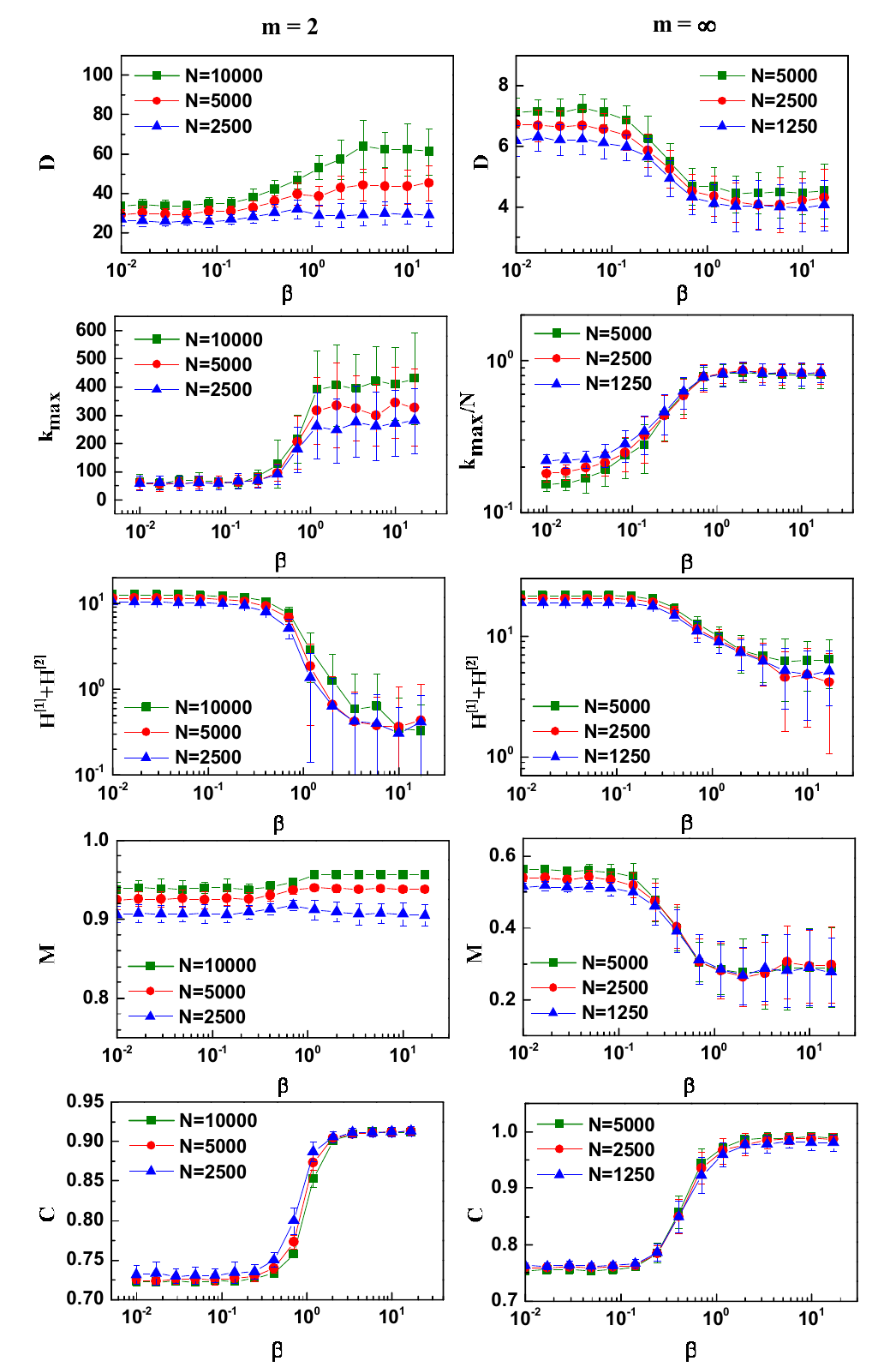}
\caption{(Color online)The maximal distance $D$ from the initial triangle, the maximal degree $k_{max}$,  the entropy rate  $H^{[1]}+H^{[2]}$, the modularity $M$ calculated using the Louvain algorithm \cite{Louvain}, 
and the average clustering coefficient $C$, are plotted as a function of the inverse temperature $\beta$ for the Generalized Fermi-Dirac Network ($m=2$) and for the Generalized Bose-Einstein Network ($m=\infty$)  with $p=0.9$. 
The networks have links with energies following a Poisson distribution $g(\omega)$ with average $c=10$. The data are reported for networks of size $N=10,000$  (averaged $30$ times), $N=5000$ (averaged $60$ times) 
$N=2500$ (averaged $90$ times) and $N=1250$ (averaged $120$ times). }
\label{transition_p0b}
\end{figure*}

\begin{figure*}
\includegraphics[width=1.5\columnwidth]{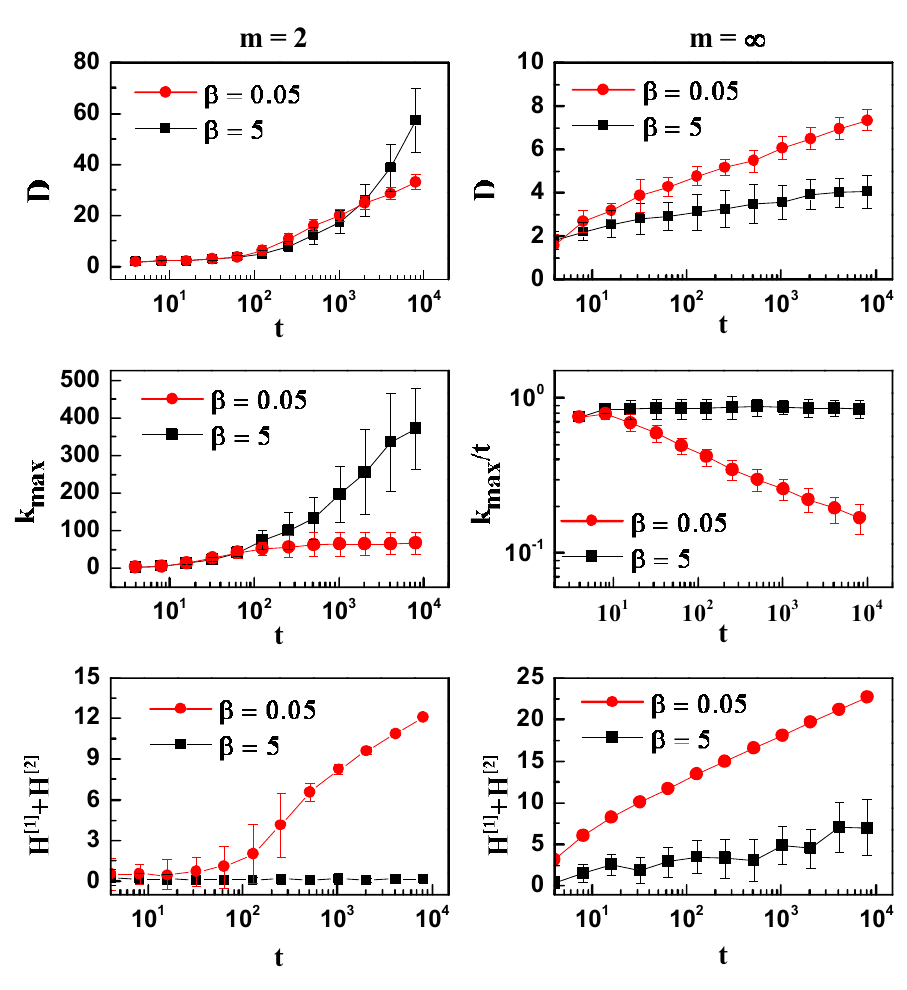}
	\caption{(Color online) The maximal distance $D$ from the initial triangle, the maximal degree $k_{max}$, and the entropy rate  $H^{[1]}+H^{[2]}$ are plotted as a function of time $t$ for the Generalized Fermi-Dirac Network ($m=2$) and for the 
	Generalized Bose-Einstein Network ($m=\infty$) with $p=0.9$. The inverse temperatures are  $\beta=0.05$ and $\beta=5$ respectively, below and above the phase transitions.	
	The networks have nodes with energies following a Poisson distribution $g(\omega)$ with average $c=10$. The data are averaged $20$ times. }
	\label{transition_time_p0b}
\end{figure*}
\begin{figure}
	\includegraphics[width=0.95\columnwidth]{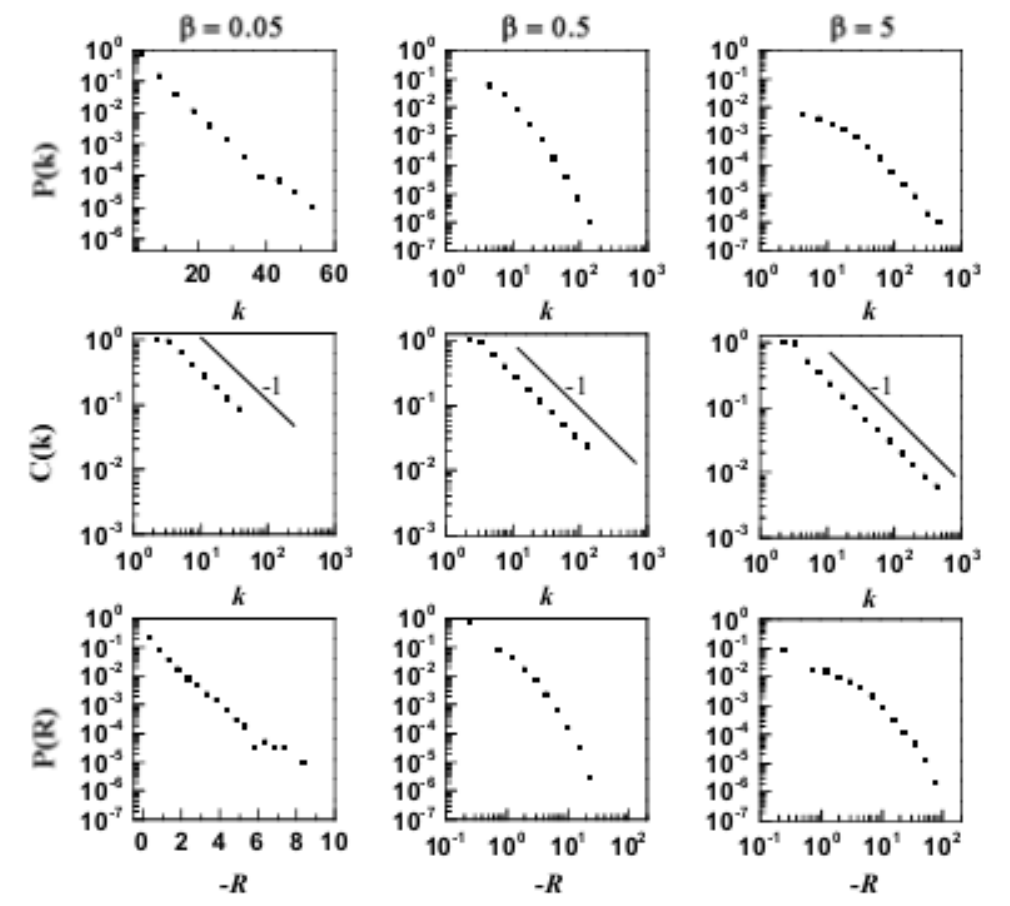}
	\caption{Structural and geometrical properties of the Generalized Fermi-Dirac Network as a function of the inverse temperature $\beta$ for single network realizations of size $N=10^5$. 
	The degree distribution $P(k)$, the average clustering coefficient $C(k)$ 
	of nodes of degree $k$ and  the distribution of the  curvatures $P(R)$ are plotted for $\beta=0.05,0.5,5$ and $p=0.9$. The networks have nodes with energies following a Poisson distribution 
	$g(\omega)$ with average $c=10$. For low values of $\beta$, i.e. $\beta<\beta_c$, both 
	$P(k)$ 
	and 
	$P(R)$ are exponential, while for large values of $\beta$, i.e. $\beta>\beta_c$, they become power-law. 
	The average clustering coefficient $C(k)$ of nodes of degree $k$ 
	$C(k)$ always goes like $C(k)\propto k^{-1}$. }	
	\label{figuresimm2b}
\end{figure}
\begin{figure}
	\includegraphics[width=0.95\columnwidth]{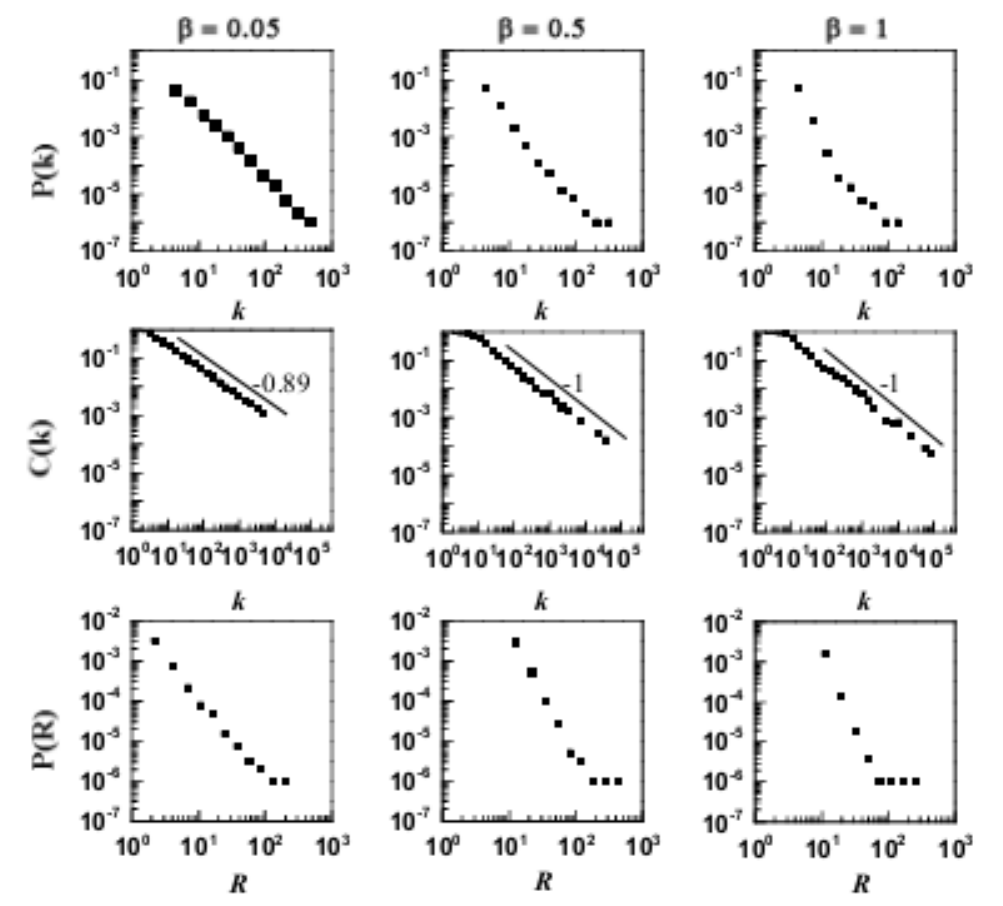}
	\caption{Structural and geometrical properties of the Generalized Bose-Einstein Network as a function of the inverse temperature $\beta$ for single network realizations of size $N=10^5$. 
	The degree distribution $P(k)$, the average clustering coefficient 
	$C(k)$ of nodes of degree $k$ and  the distribution of the curvatures $P(R)$ are plotted for $\beta=0.05,0.5,1$ and $p=0.9$. The networks have nodes with energies following a Poisson 
	distribution $g(\omega)$ with average $c=10$. For low values of $\beta$,  i.e. $\beta<\beta_c$, both 
	$P(k)$ and 
	$P(R)$ are scale-free while for large values of $\beta$, i.e. $\beta>\beta_c$, these distributions become dominated by outliers. 
	The average clustering coefficient 
	$C(k)$ 
	of nodes of degree $k$ 
	always goes like $C(k)\propto k^{-\alpha}$ with $\alpha\leq1$.}	
	\label{figuresimm1b}
\end{figure}

\subsection{Phase transition in the generalized geometric networks}
Except for the case $m=2$ which is planar, the generalized geometric network model with $p>0$ is not planar, 
and we  have 
\bea
\lim_{t\to \infty}\frac{\chi}{t}>0.
\eea
In fact,  if process $(b)$ occurs, we add zero new nodes, one new link, and we have the possibility to add more than one triangle if $m>2$. Therefore 
we have that at any given time 
\bea
\Avg{\Delta \chi}=\Avg{\chi(t)-\chi(t-1)}>0.
\eea
For these networks we extend the definition of the  curvature defined for planar graphs \cite{Keller1,Keller2} and we take the  curvature $R_i$ associated to each node $i$ of the network given by 
\bea
R_i=1-\frac{k_i}{2}+\frac{T_i}{3},
\eea
which is the same definition as in Eq. (\ref{networkcurvature}) apart from that there is no simple relation between $k_i$ and $T_i$.
Here we focus in particular on the cases $m=2$ and $m=\infty$ and we show numerical evidence that a phase transition might occur in these networks.
Specifically, we consider the case  in which $\omega$ can only take integer values and  the distribution $g(\omega)$ is Poisson with average $c$, i.e. Eq. (\ref{Poissondistribution}),
and node and link energies are related by Eq. (\ref{linearenergy}).

As a function of $\beta$ we observe a phase transition both in the case of the Generalized Fermi-Dirac Network 
and in the case of the Generalized Bose-Einstein Network. 
For $\beta>\beta_c$ the structure of the network and its geometry change drastically as it can been seen already from the visualizations of the networks (Figure $\ref{visualization_m2b}$ for the Generalized Fermi-Dirac Network
and Figure $\ref{visualization_m1b}$ for the Generalized Bose-Einstein Network).
The transition is characterized by a different scaling of the entropy rate $H^{[1]}(t)+H^{[2]}(t)$ below and above the transition. For $\beta<\beta_c$, $H^{[1]}(t)+H^{[2]}(t)$ 
increases with time as 
$H^{[1]}(t)+H^{[1]}(t)\simeq \ln(t)$. For $\beta>\beta_c$ instead,  $H^{[1]}(t)+H^{[2]}(t)={\cal O}(1)$ fluctuates widely during the network evolution.
Here we discuss in detail the consequences of this phase transition in the Generalized Fermi-Dirac Network 
and in the Generalized Bose-Einstein Network.
In Figure $\ref{transition_p0b}$ we show major geometrical and structural properties of the network as a function of the inverse temperature $\beta$ across the phase transitions. 
In particular we display the maximal shortest (hopping) distance from the initial triangle $D$, the maximum degree $k_{max}$ of the network, the entropy rate $H^{[1]}+H^{[2]}$, the modularity $M$ 
calculated using the Louvain algorithm \cite{Louvain}  and the average clustering coefficient $C$ across the phase transitions.
In Figure $\ref{transition_time_p0b}$ we show major geometrical and structural properties of the network as a function of time for given values of 
$\beta$ 
below and above the transition for $p=0.9$.
Finally in Figures $\ref{figuresimm2b}$ and $\ref{figuresimm1b}$ we show the degree distribution $P(k)$,  the average clustering coefficient $C(k)$ of nodes of degree $k$,  and the distribution 
of the curvature $P(R)$, for  the Generalized Fermi-Dirac Network and for the Generalized Bose-Einstein Network below and above the phase transition.

In  the Generalized Fermi-Dirac Network, for  $\beta<\beta_c$, 
$D$  grows logarithmically with time,( i.e.
the network is small-world)
and  the degree distribution 
$P(k)$ is exponential.
For $\beta>\beta_c$, $D$  
grows as a power-law with time,
(i.e. the network is not any more small-world)
$k_{max}$ increases significantly
and $P(k)$ 
follows a power-law.
For every value of $\beta$, the network has high modularity $M$ 
and a hierarchical structure \cite{Ravasz} with an average clustering coefficient $C(k)$ of nodes of degree $k$ decaying as $C(k)\simeq k^{-\alpha}$ and $\alpha=1$.
The curvature distribution  $P(R)$ has a negative tail.\\

In  the Generalized Bose-Einstein Network, for $\beta<\beta_c$,
$k_{max}$ scales sub-linearly with the network size, 
$D$ increases logarithmically with the network size, (i.e. the network is small-world) and
the degree distribution 
$P(k)$ is scale-free,
the network has high modularity $M$ 
and $P(R)$ 
has a scale-free positive tail.
For $\beta>\beta_c$, 
$k_{max}$ scales linearly with the network size, (i.e. the largest node is linked to a finite fraction of the links), 
$D$ decreases significantly, 
$P(k)$ is dominated by outlier hubs,
the network has low modularity $M$ 
and $P(R)$ has a positive tail dominated by outliers.
For every value of $\beta$, the network has a hierarchical structure \cite{Ravasz} with an average clustering coefficient $C(k)$ of nodes of degree $k$ decaying as $C(k)\simeq k^{-\alpha}$ and $\alpha\leq 1$.

\section{Generalized Quantum Network Evolution}
\subsection{The evolution}
Here we consider a generalized quantum network state evolution corresponding to the generalized geometric network evolution using a similar approach as in the case $p=0$.
We start from an initial condition given by 
\bea
\ket{\psi_N(1)} =\frac{1}{\sqrt{{\cal Z}(1)}}\sum_{\omega_1,\omega_2,\omega_3} \left[\prod_{i=1,3}\rho(\omega_i)b^{\dag}_i(\omega_i)\right]c^{\dag}_{12}c^{\dag}_{23}c^{\dag}_{13}\ket{0},
\eea
where ${\cal Z}(1)$ enforces the normalization condition $\Avg{\psi_N(1)|\psi_N(1)}$
The evolution of the quantum network state is a non-equilibrium Markovian dynamics obtained applying the unitary operator $U_t$ to the state $\ket{\psi_N(t-1)}$. This dynamics and the operator $U_t$ are  defined as in the following, 
\bea
&&\ket{\psi_N(t)}=U_t\ket{\psi_N(t-1)}\nonumber \\
&&=\sqrt{\frac{{\cal Z}(t-1)}{{\cal Z}(t)}}\left[\sqrt{1-p}+\sqrt{p}\sum_{i^{\prime},j^{\prime},r^{\prime}|i^{\prime}<r^{\prime}}e^{-\beta(\epsilon_{i^{\prime}j^{\prime}}+\epsilon_{j^{\prime}r^{\prime}})/2} \right.\nonumber \\
&&\times \left.c^{\dag}_{i^{\prime}r^{\prime}}\left(\prod_{s}h^{\dag}_{i^{\prime}s}h^{\dag}_{sr^{\prime}}\right)c^{\dag}_{j^{\prime}r^{\prime}}c_{j^{\prime}r^{\prime}}
c^{\dag}_{i^{\prime}j^{\prime}}c_{i^{\prime}j^{\prime}}\right]\nonumber \\
&&\times\sum_{\omega_{t+2}}\sum_{i,j|i<j}\sqrt{g(\omega_{t+2})}e^{-\beta\epsilon_{ij}/2}\nonumber \\
&&\times b^{\dag}_{t+2}(\omega_{t+2})c^{\dag}_{(t+2)i}c^{\dag}_{(t+2)j}h^{\dag}_{ij}c^{\dag}_{ij}c_{ij}\ket{\psi_N(t-1)}
\label{EG}\eea
where ${\cal Z}(t)$ is fixed by the normalization condition $\Avg{\psi_N(t)|\psi_N(t)}=1$.
Moreover  the quantum operator $h^{\dag}_{ij}$, in Eq. $(\ref{EG})$  depends on the type of the quantum network state, i.e. 
\bea
h^{\dag}_{ij}=\left\{\begin{array}{ll} d^{\dag}_{ij} & \begin{array}{c}\mbox{ Generalized Fermi-Dirac}\nonumber \\ \mbox{ Quantum Network State}\end{array}\nonumber \\
& \nonumber \\
\tilde{d}^{\dag}_{ij} & \begin{array}{c}\mbox{ Generalized Bose-Einstein}\nonumber \\ \mbox{ Quantum Network State}\end{array}\end{array}\right. .
\eea
With a long but straightforward calculation,  following the same steps as for the original Fermi-Dirac Network state and the original Bose-Einstein network state,  
it is possible to show that the normalization constant ${\cal Z}(t)$ can be interpreted as a path integral over the paths corresponding to the evolution of the 
Generalized Growing Geometric  Networks 
allowing us to obtain the Generalized Fermi-Dirac Network for $m=2$ and the  Generalized Bose-Einstein Network for $m=\infty$. 

\section{Dual networks and Connection between the Fermi-Dirac Quantum Network and Spin Networks}

\begin{figure}
	\includegraphics[width=0.95\columnwidth]{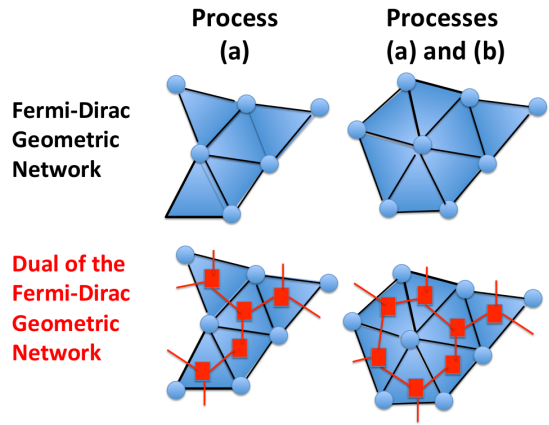}
	\caption{(Color online) Fermi-Dirac Network and its dual network. In the case in which only process (a) takes place the dual network is a tree. In the case in which also process (b) takes place the dual network contains loops. 
	In this case we have explicitly drawn the triangles forming the underlying simplicial complex of the Fermi-Dirac networks. The nodes of the dual network are plotted as red squares and correspond to the triangles of 
	the Fermi-Dirac network. The links $\ell$ of the dual network correspond to the links $(i,j)$ between nodes $i$ and $j$ of the Fermi-Dirac network and are associated to a spin variable $J_{\ell}=J_{ij}=\epsilon_{ij}/2$. The links of the dual are indicated with red lines and can be saturated (if they join two nodes of the dual network) or unsaturated if they connect only to a single node of the dual network.}	
	\label{figuredual}
\end{figure}

Starting from the Fermi-Dirac Quantum network we can construct the dual network
by mapping each triangle of the original network to a node of the dual network and every link of the original network to a link in the dual network (see Figure $\ref{figuredual}$).

Each node of the dual  of the Fermi-Dirac Network has degree $3$. 
A link of the dual network can be saturated or unsaturated.
A link of the dual network is saturated if it connects two nodes of the dual network. This happens if and only if the two corresponding triangles of the original network are glued together (i.e. they have a common link).
A link of the dual network is unsaturated if it  does not connect two nodes. This happens if a triangle of the original network has an unsaturated link. 

As the Fermi-Dirac network grows, also the dual network grows.
In the  original Fermi-Dirac Network  only process (a) takes place, therefore  the dual network is a tree. If instead one considers  the Generalized Fermi-Dirac Network, in which also process (b) 
takes place, the dual network contains loops (see Figure $\ref{figuredual}$).

In the case in which the energies of the nodes take only integer values we can interpret the dual network as  a spin network \cite{spinnet1,spinnet2,Rovelli}.
In fact we can  associate to the link $\ell=(i,j)$ of the dual network  the   half-integer spin $J_{ij}$ given by Eq. $(\ref{spin})$ 
satisfying the Clebsch-Gordon conditions given by Eq. $(\ref{triangulard})$ at each node of the dual network.
Nevertheless, we  note here that the quantum evolution of the proposed Fermi-Dirac network is a non-equilibrium dynamics  and not an equilibrium one  as usually assumed in the context of spin networks.

In the case of the Bose-Einstein Networks it is also possible to construct the dual network following a similar procedure used for constructing the dual of the Fermi-Dirac Network, but in this case 
the dual will not be regular, since each triangle in the Bose-Einstein network model can be linked to an arbitrarily large number of other triangles incident to the links at its boundary.

 It is to mention that in the literature  of quantum gravity spin networks with a causal structure have been proposed and  are the so called energetic causal sets \cite{Smolin1,Smolin2,Smolin3}. 
It will be therefore interesting to explore further the connections between the Fermi-Dirac Networks and energetic causal sets. 

Our framework is instead quite far from the spin networks used in Loop Quantum Gravity.  Notably in complex quantum network geometries we do not make use of intertwines, and the network does not have relevant simple symmetries. 
 
\section{Relation to triangulations, foams, and planar graphs}

Planar complex networks have already been studied in the literature in several contexts \cite{Planar,Doro_link,det,Apollonian, Aste2,Aste3, Eckmann}, including the study of glass and foams, and planar complex networks.

The model most closely related to our model is the scale-free random network constructed by adding randomly triangles to links \cite{Doro_link}. Nevertheless, our model does not reduce to this model for any value of the parameters. In fact also the Bose-Einstein Network for $\beta=0$ is not equivalent to this model.
The difference is that in the Bose-Einstein network at $\beta=0$, each link is not chosen randomly, but proportionally to the number of triangles  already incident to it (i.e. $1+n_{ij}$), according to a kind of "preferential attachment" to the link.

Other  planar scale-free network models are the pseudo-fractal scale-free network \cite{det} and the Apollonian networks \cite{Apollonian}.
These network models are deterministic and yield scale-free networks with given power-law exponent, while the  Bose-Einstein Networks are stochastic  planar scale-free networks whose power-law exponent depends on the distribution $g(\omega)$ and on the value of the inverse temperature $\beta$.

Other models for complex networks embedded into surfaces have been recently proposed \cite{Aste2} extending  approaches used already in the study of glasses and foams \cite{Aste3,Eckmann}.
In \cite{Aste2} maximal embedded graphs have been characterized using a Monte Carlo algorithm  determined by an Hamiltonian  which is a function of the degree of the nodes.
This dynamics can explore networks embedded in surfaces of different genus, and displays a dynamical slowing down as a function of the inverse temperature of the Monte Carlo algorithm. Therefore in these simulations the ground state, ordered network is not observed, similarly to what has been  observed in models of glass and foam dynamics \cite{Eckmann,Aste3}.
This approach is very different from the one proposed  in the present article, although the focus is always the characterization of the network geometry.
For  example the phase transitions present in the Quantum Complex Network Geometries are very different from the one observed in \cite{Aste2,Aste3,Eckmann}. In fact in the Complex Quantum Network Geometries the observed  
phase transitions are non-equilibrium phase transitions,  they are determined by the quenched disorder in the network. Instead, in  \cite{Aste2,Aste3,Eckmann} the network has an equilibrium Hamiltonian dynamics, the ground state is well defined but there is a dynamical slow down.  
Moreover, in Complex Quantum Network Geometries the observed phase transitions can change the metric properties of the networks, which can go from a 
small-world network to a network with large diameter  without changing its Euler characteristic, as  in the case of the phase transition observed for Fermi-Dirac networks. Instead in  \cite{Aste2}, changes from small world networks to  networks with large diameter occur as a function of the Euler characteristic of the network.

\section{Conclusions}
In this work we have proposed to study a geometrical network evolution based on growing simplicial complexes of dimension $2$, i.e. simplicial complexes formed by triangles.
The network evolution describes the evolution of a quantum network state defined by a path integral.
The quantum network states are characterized by a set of quantum occupation numbers that can be mapped to the existence of nodes, links and triangles incident to the links of the geometric  network.
In particular we distinguish between Fermi-Dirac network states in which the  incident triangles quantum occupation number can  take only values $0,1$, and the Bose-Einstein quantum network state where it can take  
any possible integer value $0,1,2,\ldots$
The Fermi-Dirac Network describes the evolution of the quantum Fermi-Dirac network state and the Bose-Einstein Network describes the evolution of the quantum Bose-Einstein network state. 
The average of the number of triangles  exceeding one and incident to  links of energy $\epsilon$  follows the Fermi-Dirac and the Bose-Einstein statistics in the Fermi-Dirac and in 
the Bose-Einstein Networks respectively.
The Fermi-Dirac and the Bose-Einstein Networks are complex networks, including the small-world property, high clustering coefficient, exponential and scale-free degree distributions, and high modularity.
The proposed network models have an emergent random geometry, since their Euler characteristic can go from $\chi=1$ (planar networks) to $\chi\propto N$ where $N$ are the number of nodes in the network, and can have a non-trivial 
distribution of the curvature $P(R)$.
Moreover these networks can have a phase transition as a function of the external parameter $\beta$ called the inverse temperature.
For $\beta>\beta_c$ the Fermi-Dirac network is not any more small-world and the diameter $D$ grows as a power-law of the number of nodes $N$.
For $\beta>\beta_c$ the Bose-Einstein Network undergoes a Bose-Einstein condensation, and one link acquires a finite fraction of triangles, therefore the two nodes at the two ends of the link have a finite fraction of all the links.

We believe that this model can be used to explore further the relation of complex geometries with quantum mechanics.
  
In the future, we plan to explore further the geometrical network model by characterizing the networks with general values of the parameter $m$, defining  the corresponding quantum state evolution 
and characterizing further their topological and geometrical properties.
Moreover,  we plan to study how quantum dynamical processes \cite{Ising,BH,BHJ,Jesus1,Garneronegoogle} can be affected by the structure of these networks.
Finally we plan to consider equilibrium network models describing the underlying structure of simplicial complexes  and to explore the possibility to observe further structural 
phase transitions in geometrical complex networks.\\


\begin{thebibliography}{99}
\bibitem{note} 
Note that throughout the paper we give physically inspired names to mathematical quantities describing the complex quantum network evolution, but no indication is given to an experimental  implementation of  the complex quantum network geometries or of the quantum  network states.
\end{thebibliography}
\end{document}